# Gaussian Integral Method for Void Fraction

Alireza Kianimoqadam[1]

Justin Lapp[1]

1. University of Maine, Department of Mechanical Engineering, 75 Long Road, Orono, ME, 04469, USA

## Abstract

A novel method, the Gaussian Integral Method (GIM), is presented for calculating void fractions in Computational Fluid Dynamics–Discrete Element Method (CFD-DEM) simulations. GIM is versatile and applicable to various grid types, including structured and unstructured polyhedral meshes, without requiring special boundary treatments. An optimization technique is introduced to make GIM independent of grid resolution and type. The method is validated against experimental data from a fluidized bed, demonstrating that GIM produces realistic simulations closely resembling experimental observations. Additionally, unstructured polyhedral grids using GIM outperform structured grids of equivalent resolution, yielding results more aligned with experimental data. The gradient of the void fraction is computed in the CFD solver and utilized in the DEM solver for precise estimation at particle locations. Overall, GIM provides an effective solution for void fraction calculations in particulate media simulations with complex geometries, enhancing the accuracy and applicability of CFD-DEM simulations in industrial applications.

## Introduction

Multiphase systems, where particles interact with fluids, are essential in industries like chemical, metallurgical, energy, food-processing, pharmaceutical, and mineral processing [1-5]. These systems impact over half of industrial products. Traditionally, understanding them relied on empirical methods. However, advances in computer science and computational techniques have introduced powerful tools like Computational Fluid Dynamics and the Discrete Element Method (CFD-DEM) [6, 7]. These methods are vital for analyzing these systems, tracking particle motion with a Lagrangian approach, and fluid dynamics from an Eulerian perspective. They also address the physics of particle interactions and the transfer of momentum, heat, and mass [7-9].

Despite the progress, there are still challenges in coupling CFD and DEM efficiently, particularly when dealing with complex industrial geometries and various CFD meshes. Implementations range from standalone applications, as highlighted by Garg et al., to integrated ones like the coupling of OpenFOAM with LIGGGHTS, as cited by Kloss et al [10-13]. In the previous work of the authors, a GPU-based DEM solver was developed and optimized which accelerated the computational speed by a factor of 2.57 compared to its predecessor [14, 15]. Furthermore, polyhedral meshes were employed for CFD-DEM simulations relating to complex geometries.

The most challenging issue with polyhedral meshes is accurately calculating the void fraction and distributing the fluid-phase source terms. The void fraction—the proportion of fluid volume within each computational cell—is crucial because it directly affects the calculation of fluid-particle interaction forces, such as forces due to drag and pressure gradient. These forces are essential for accurately predicting how particles move within the fluid flow. Inaccurate void fraction calculations can lead to errors in these forces, resulting in incorrect momentum and energy exchanges between the fluid and particles. This not only impacts the fidelity of the simulation but can also lead to numerical instability. Polyhedral meshes, with their complex cell shapes and varying sizes, make it particularly difficult to determine how particles overlap with computational cells. This complicates the precise calculation of local void fractions and the proper distribution of source terms in the fluid equations, which are essential for capturing the physics of particle-fluid interactions accurately.

Accurately calculating the void fraction—the proportion of fluid volume within a computational cell—is crucial for reliable CFD-DEM simulations, as it directly influences the prediction of particle-fluid interactions. There are several methods available for this calculation, each with its own strengths and challenges. Analytical methods offer exact solutions but can be computationally intensive and difficult to implement, especially in complex geometries. Non-analytical approximations, such as statistical kernel methods and distributed particle volume methods, provide computational efficiency but may sacrifice some accuracy.

In previous studies on void fraction calculations, Peng et al. (2014) emphasized that accurately determining the void fraction is crucial for reliable CFD-DEM simulations of gas-solid bubbling fluidized beds, as inaccuracies can lead to significant deviations in simulation results [16]. They employed an analytical method for calculating the void fraction, which offers exact solutions and high accuracy when using structured hexahedral grids. This method meticulously accounts for the geometric overlap between particles and computational cells, ensuring precise calculations.

However, the analytical approach comes with notable challenges. It is computationally intensive and difficult to implement, especially when dealing with complex geometries. The method's reliance on detailed geometric calculations makes it less adaptable to unstructured grids commonly used in practical simulations with complex domains. Consequently, its application is limited in scenarios where the computational domain or particle geometry is not straightforward.

In addition to the limitations of analytical methods, the choice of computational cell size significantly impacts the accuracy of void fraction calculations. Volk and Ghia (2019) investigated how varying computational cell sizes affect CFD-DEM simulations, revealing that smaller cell sizes can actually lead to significant numerical errors due to increased porosity fluctuations [17]. As the cell size approaches the particle size, local porosity values become highly variable, introducing instability and inaccuracies into the simulation. They emphasized the importance of carefully selecting cell sizes to balance accuracy and computational efficiency, highlighting the limitations of traditional void fraction calculation methods when applied to very fine meshes.

These challenges with both the analytical methods and cell size selection underscore the need for improved approaches that can accurately calculate void fractions in complex geometries and across a range of cell sizes. Direct calculation methods aim to strike a balance between accuracy and computational efficiency by directly computing the overlap volume between particles and computational cells, enhancing precision while maintaining reasonable computational costs.

Direct calculation methods aim to strike a balance between accuracy and computational efficiency by directly computing the overlap volume between particles and computational cells. Peng et al. (2016) introduced a modified direct method called the Particle Meshing Method (PMM), which improves void fraction calculation by meshing particles into smaller elements. This approach enhances accuracy by more precisely accounting for particle-cell interactions while maintaining computational efficiency. PMM also exhibits the potential to handle arbitrarily complex domain and particle geometries [18].

Building on this, Wang et al. (2018) developed a direct calculation method specifically for CFD-DEM simulations of fluidized beds with large particles. They proposed a judgment criterion for particle-cell overlap and unified eight cases of particle-cell overlap into a single formula. Their method demonstrated high accuracy and reliability through extensive validation against both experimental and simulation data, effectively capturing the hydrodynamic behavior of three-dimensional fluidized beds [19].

Clarke et al. (2018) investigated various void fraction schemes for use with CFD-DEM simulations of fluidized beds, offering a comprehensive analysis of different approaches, including the Particle-Centered Method (PCM), Divided Particle Volume Methods (DPVM), Statistical Kernel Methods, and Satellite Point Methods. They evaluated these schemes based on their accuracy and computational efficiency when applied to simulations with different grid resolutions. Their study provides valuable insights into how the choice of void fraction calculation impacts simulation results, emphasizing the importance of selecting an appropriate scheme to ensure reliable and accurate simulations for various grid configurations and fluidized bed setups [20].

The Gaussian methods provide a statistical framework for calculating void fractions, which helps manage spatial inconsistencies and noise in the data. Integrating these methods into CFD-DEM simulations allows researchers to make more accurate and reliable estimations, improving the design and optimization of multiphase flow systems. Takabatake and Sakai (2020) introduced a flexible discretization technique for CFD-DEM simulations that includes thin walls, enhancing the precision of void fraction calculations in complex geometries [21]. Zhang et al. (2020) also developed a semi-resolved CFD-DEM method for thermal particulate flows, highlighting its use in fluidized beds and the importance of accurate void fraction calculations in thermal systems [22].

Xiao and Sun (2011) proposed a CFD-DEM solver that effectively calculates void fractions using an interpolation algorithm for unstructured meshes, which reduces mesh dependence. They also introduced a semi-implicit treatment for fluid-particle interaction terms, improving stability and accuracy [23].

Sun and Xiao (2015) introduced a new coarse-graining algorithm for continuum-discrete solvers. This method, which uses a transient diffusion equation, offers a flexible and straightforward solution for CFD-DEM solvers that work with various meshes. Although theoretically similar to the Gaussian kernel function method, this diffusion-based approach is more straightforward. Preliminary numerical tests showed that this new method produces smooth and consistent results even on unfavorable meshes, maintaining physical quantities accurately in theory [24]. In their future companion paper, they employed the method in OpenFOAM and LAMMPS CFD-DEM solver. However, They tested the method only for structured meshes [25]. This algorithm is a significant step forward in calculating void fractions and has been successfully integrated into a CFD-DEM solver, providing better mesh convergence than existing methods.

Peng et al. demonstrated that the particle-centered method (PCM) is neither suitable nor practical for void fraction calculations in CFD-DEM simulations, as it can cause high and unreasonable pressure fluctuations [16]. Using Wachem et al.'s case study, they observed that PCM can lead to inaccurate results because it assigns the entire volume of a particle to the cell containing the particle's center. In reality, since a particle often overlaps multiple adjacent cells, especially when the cell size is comparable to the particle size, the particle's volume is distributed among these cells. This distribution is indeed dependent on the cell-to-particle size ratio, geometries, and their relative positions. This discrepancy can result in unrealistically low void fractions in some cells, causing excessive drag forces. The high drag forces then lead to significant pressure fluctuations in the fluid phase. Precise void fraction calculations are essential for obtaining results that are close to experimental observations, as evidenced by Peng et al.'s use of an analytical approach for structured grids, which produced more stable and realistic pressure signals.

This study introduces the Gaussian Integral Method (GIM) for calculating void fraction. Validated by experimental data, GIM demonstrates effectiveness across different grid types, including structured and unstructured polyhedral meshes. Polyhedral grids offer a compromise between the precision of structured grids in fluid field gradient calculations and the adaptability of tetrahedral grids for complex geometries. The use of polyhedral cells in CFD-DEM simulations enhances the accuracy of fluid field gradients due to their larger number of neighboring cells compared to tetrahedral grids. This increased connectivity allows for more precise and stable simulations, particularly in intricate and detailed geometrical domains.

GIM enhances void fraction calculations by utilizing the statistical properties and smooth distribution techniques inherent in Gaussian functions. In this method, Gaussian smoothing functions are applied to both the particles and the surrounding CFD cells, with each function scaled according to the volume of its respective object. This scaling ensures that an object's influence in the calculation is proportional to its size. The effectiveness of this method depends on the cell-to-particle size ratio as well as the geometry and relative positioning of the particles and cells. While the GIM approximates these geometries using scaled Gaussian functions based on volume ratios, it is important to note that the accuracy of the void fraction distribution can vary with changes in the volume ratio and geometric arrangement. This dependency underlines the method's sensitivity to the mesh configuration and the importance of considering these factors in simulations.

The core concept involves determining the overlap between the Gaussian functions of particles and cells. The intersection of a particle's Gaussian function with that of a cell represents the fraction of the particle's volume associated with that specific cell. By calculating the area shared under these smoothing functions—the shared integral—the volume ratio of the particle belonging to the cell is determined.

To limit the computational domain and ensure efficiency, the smoothing functions are bounded within a specified radius based on the cubic root of the object's volume. This approach naturally restricts the influence of particles and cells without requiring imaging techniques or special boundary treatments [24, 26].

By summing the calculated volume ratios (GIM ratios) for each cell surrounding a particle, the particle's total volume is accurately distributed among its neighboring cells, enhancing the precision of the void fraction calculation.

Additionally, the CFD solver computes the gradient of the void fraction at each time step. This gradient information is utilized in the DEM solver to precisely estimate the void fraction at the location of each particle, further improving the accuracy of particle-fluid interaction calculations.

Overall, the Gaussian Integral Method offers a versatile and precise solution for void fraction calculations in CFD-DEM simulations. Its applicability to various grid types—including structured and unstructured polyhedral meshes—and its ability to prevent artificial volume accumulation make it a valuable tool for simulations involving complex geometries common in industrial applications.

## Mathematical Model

The CFD-DEM approach is used to study the dynamics of granular media within fluid flows. In this model, the fluid is treated as a continuum, with its mass conservation and dynamics solved using the finite volume method. The DEM solver handles the granular media, addressing interparticle, particle-wall, and particle-fluid interactions. Interaction between the fluid and solid phases is managed through an interface that transfers calculated fluid velocities, pressures, and field gradients from the CFD solver to the DEM solver. Conversely, the fluid field momentum source terms and void fractions computed by the DEM solver are transferred back to the CFD solver. This bidirectional exchange, known as two-way coupling, is employed in this study. The solver for granular media is a GPU-based DEM solver that was introduced and validated in our previous publications [14, 27].

## Governing Equations

The governing equations of the solid phase are described as follows. The linear momentum of the particles based on Newton's second law is defined by

$$m_i \frac{d\boldsymbol{v}_i}{dt} = m_i \boldsymbol{g} + \sum_j \boldsymbol{F}_{n,ij} + \sum_j \boldsymbol{F}_{t,ij} + \boldsymbol{F}_{d,i} - V_{p,i} \nabla P \tag{1}$$

Where $\boldsymbol{v}_i$ is the velocity vector, $V_{pi}$ the volume, and $m_i$ the mass of particle $\boldsymbol{i}$. $\boldsymbol{g}$ is the gravitational acceleration vector. $\boldsymbol{F}_{n,ij}$ and $\boldsymbol{F}_{t,ij}$ are the normal and tangential contact forces between particle $\boldsymbol{i}$ and particle $\boldsymbol{j}$ respectively. $\boldsymbol{F}_d$ and $\nabla P$ are the drag force and the local pressure gradient of the fluid phase.

The torques acting on particles because of tangential forces are described as follows [28].

$$I_i \frac{d\boldsymbol{\omega}_i}{dt} = \sum_j \boldsymbol{r}_i \times \boldsymbol{F}_{t,ij} \tag{2}$$

Where $I_i$ and $\boldsymbol{\omega}_i$ are the moment of inertia of angular velocity of particle $\boldsymbol{i}$ respectively. $\boldsymbol{r}_i$ denotes the vector that goes from the particle's center of mass to the collision contact point.

The collision forces (normal and tangential) exerted on particle $\boldsymbol{i}$ are calculated based on Hertz -Mindlin contact theory [29, 30].

The fluid phase mathematical model is based on the volume-averaged continuity and Navier-Stokes equations of porous media, described as follows [31].

$$\frac{\partial(\alpha_f \rho_f)}{\partial t} + \nabla \cdot \left(\alpha_f \rho_f \boldsymbol{v}_f\right) = 0 \tag{3}$$

$$\frac{\partial(\alpha_f \rho_f \boldsymbol{v}_f)}{\partial t} + \nabla \cdot \left(\alpha_f \rho_f \boldsymbol{v}_f \boldsymbol{v}_f^T\right) = \nabla \cdot \left(\alpha_f \boldsymbol{\tau}_f\right) - \alpha_f \nabla P + \alpha_f \rho_f \boldsymbol{g} + \boldsymbol{S}_p \tag{4}$$

Where $\boldsymbol{\alpha}_f$, $\boldsymbol{\rho}_f$, and $\boldsymbol{v}_f$ are the local void fraction, density, and velocity of the fluid phase respectively. $\boldsymbol{S}_p$ is the momentum source term that results from the local solid phase (particles in a CFD cell) and is exerted on the fluid phase. $\boldsymbol{\tau}_f$ is the viscous stress tensor. The viscous stress tensor for a Newtonian incompressible fluid is described as follows.

$$\boldsymbol{\tau}_f = \mu_f \left[\nabla \boldsymbol{v}_f + \left(\nabla \boldsymbol{v}_f\right)^T - \frac{2}{3}\left(\nabla \cdot \boldsymbol{v}_f\right)\mathbf{I}\right] \tag{5}$$

Where $\mu_f$ and $\mathbf{I}$ are the shear viscosity of the fluid and the identity matrix respectively.

The momentum source terms exerted on the fluid phase are due to the sum of forces exerted on the local solid phase (particles in a CFD cell) by the fluid phase in the reverse direction. The drag force and the force due to the fluid phase pressure gradient are the fluid forces acting on the solid phase.

The Beetstra-van der Hoef-Kuipers (BVK) drag model is applied to particles in this study [32]. BVK correlation is derived from lattice-Boltzmann simulations. The BVK drag model is described as follows.

$$\boldsymbol{F}_d = \frac{V_p \beta_{BVK} (\boldsymbol{v}_f - \boldsymbol{v}_p)}{\alpha_s} \tag{6}$$

Where $\beta_{BVK}$ is the inter-phase momentum exchange coefficient that is derived by the following equations. $\boldsymbol{v}_f$ is the fluid velocity at the particle location which is estimated by the local fluid velocity gradient. $\alpha_s$ and $Re_d$ are the solid fraction and the Reynolds number respectively which are defined as follows.

$$\alpha_s = 1 - \alpha_f \tag{7}$$

$$Re_d = \frac{\alpha_f \rho_f |\boldsymbol{v}_f - \boldsymbol{v}_p| d_p}{\mu_f} \tag{8}$$

Where $d_p$is the diameter of a particle.

The drag force exerted on a particle in porous media needs to be normalized by the Stokes-Einstein relation [33].

$$\mathcal{F}(\alpha_s, Re_d) = \frac{|\boldsymbol{F}_d|}{3\pi \mu_f d_p U_f} = \frac{{d_p}^2 \beta_{BVK}}{18 \mu_f \alpha_s \alpha_f} \tag{9}$$

$\mathcal{F}(\alpha_s, Re_d)$ is the BVK normalized drag force for monodisperse systems, which is defined in the following correlation.

$$\mathcal{F}(\alpha_s, Re_d) = \frac{10\alpha_s}{{\alpha_f}^2} + {\alpha_f}^2\left(1 + 1.5\sqrt{\alpha_s}\right) + \frac{0.413 Re_d}{24{\alpha_f}^2}\left[\frac{{\alpha_f}^{-1} + 3\alpha_s\alpha_f + 8.4{Re_d}^{-0.343}}{1 + 10^{3\alpha_s}{Re_d}^{-\frac{(1+4\alpha_s)}{2}}}\right] \tag{10}$$

$U_f$ is defined as the superficial fluid velocity of the domain. After applying the Stokes-Einstein normalization, $\beta_{BVK}$ is defined as follows.

$$\beta_{BVK} = \frac{18\mu_f \alpha_s \alpha_f}{{d_p}^2}\mathcal{F}(\alpha_s, Re_d) \tag{11}$$

The momentum source term of a CFD cell is derived by the following equation.

$$\boldsymbol{S}_p = \frac{-1}{V_{cell}} \sum_{i=1}^{N_p} \boldsymbol{F}_{d,i} \qquad (12)$$

Where $V_{cell}$ and $N_p$ are the volume of a cell and the number of particles inside the cell.

# Implementation of GIM for Void Fraction

The Gaussian Integral Method (GIM) for void fraction calculation employs Gaussian distribution functions. Instead of applying the Gaussian function solely to particles, additional Gaussian functions are assigned to the cells surrounding each particle. These particle and cell Gaussian functions intersect at no more than two points. The integral shared by both functions represents the volume ratio of the particle that belongs to a specific cell. To account for the relative sizes of particles and cells, the Gaussian functions are scaled according to their respective volumes.

The Gaussian Integral Method (GIM) for void fraction calculation employs Gaussian distribution functions. Instead of applying the Gaussian function solely to particles, Gaussian functions are also assigned to the cells surrounding each particle. These particle and cell Gaussian functions intersect at specific points, and the shared integral represents the fraction of the particle's volume associated with a particular cell. To account for the relative volumes of particles and cells, the Gaussian functions are scaled according to their respective sizes.

To ensure that GIM operates effectively across various cell geometries and maintains robustness, a tuning method is employed for the standard deviation of the Gaussian functions. This tuning adjusts the standard deviation to accurately represent the volumes of particles and cells, thereby preventing unrealistic void fractions. Consequently, the accumulation of particle volume in boundary cells, which can occur if the standard deviation is too large, is avoided. By tuning the standard deviation, GIM is established as a universal method suitable for any type of geometry, ensuring both robustness and the prevention of unrealistic void fraction calculations.

Additionally, the validation and accuracy of the method are thoroughly investigated in this study, demonstrating its reliability and effectiveness. As a result, the Gaussian Integral Method eliminates the need for imaging or other boundary treatments, providing a dependable solution for void fraction calculations in particulate media simulations with complex geometries [24, 26].

A schematic illustration of the GIM for unstructured grids is shown in Figure 1. In this method, each particle and each cell is represented by a scaled Gaussian function. It's important to note that the integrals themselves do not directly represent the portion of the particle's volume. Instead, the **ratio of the shared integral** between a particle's Gaussian function and a cell's Gaussian function to the **sum of all shared integrals for that particle** represents the fraction of the particle's volume that belongs to that specific cell. By appropriately tuning the standard deviation of the Gaussian functions, these ratios can effectively capture the geometrical relationships between particles and cells.

If the standard deviation is set too large, the Gaussian functions become too spread out, causing the particle's volume to be overly smoothed across neighboring cells. This results in an unrealistic distribution of porosity, where fine details of the particle-cell interactions are lost. Conversely, if the standard deviation is too small, the Gaussian functions become too narrow, leading to a noisy porosity field with abrupt changes between neighboring cells. Therefore, selecting an appropriate standard deviation is crucial to accurately represent the void fraction distribution and maintain numerical stability.

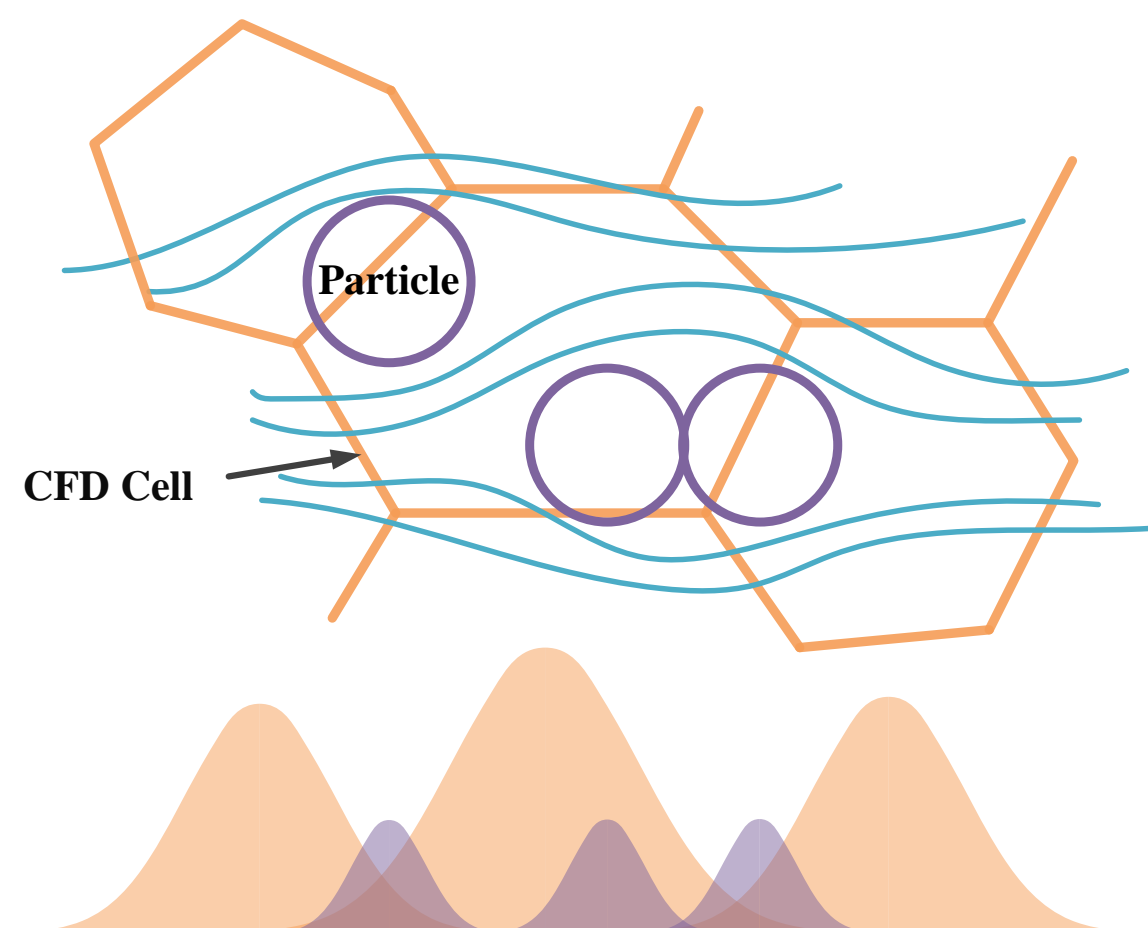

Figure 1. Schematic illustration of the Gaussian Integral Method for unstructured grid.

$$\sigma_c = \phi \left(\frac{3}{4\pi} V_c\right)^{1/3} \tag{13}$$

$$\sigma_p = \phi r_p \tag{14}$$

Where $\sigma_c$ and $\sigma_p$ are the deviations for the cell and particle Gaussian functions, and $\phi$ is the tunning coefficient for the deviations.

$$G_c(x) = \frac{V_c}{\sqrt{2\pi\sigma_c^2}} e^{\frac{-(x-\mu_c)^2}{2\sigma_c^2}} \tag{15}$$

$$G_p(x) = \frac{V_p}{\sqrt{2\pi\sigma_p^2}} e^{\frac{-x^2}{2\sigma_p^2}} \tag{16}$$

Where $G_c(x)$ and $G_p(x)$ are the Gaussian functions of cell and particle and $\mu_c$ is the cell-particle centers' distance. The intersections of $G_c(x)$ and $G_p(x)$ are determined by the following equation.

$$\mathcal{X}_{Max}, \mathcal{X}_{min} = \frac{-\sigma_p^2\mu_c \pm \sqrt{\sigma_p^4\mu_c^2 + \sigma_p^2\left[\mu_c^2 + 2\sigma_c^2 \ln\left(\frac{\sigma_c V_p}{\sigma_p V_c}\right)\right](\sigma_c^2 - \sigma_p^2)}}{\sigma_c^2 - \sigma_p^2} \tag{17}$$

If the integral boundaries ($\mathcal{X}_{Max}$ and $\mathcal{X}_{min}$) are real numbers

$$V_{p,i} = \int_{-\infty}^{x_{min}} G_p(x)dx + \int_{x_{min}}^{x_{Max}} G_c(x)dx + \int_{x_{Max}}^{+\infty} G_p(x)dx = \frac{V_p}{2}\left[2 + \text{erf}\left(\frac{x_{min}}{\sqrt{2\sigma_p^2}}\right) - \text{erf}\left(\frac{x_{Max}}{\sqrt{2\sigma_p^2}}\right)\right] + \frac{V_c}{2}\left[\text{erf}\left(\frac{x_{Max}-\mu_c}{\sqrt{2\sigma_c^2}}\right) - \text{erf}\left(\frac{x_{min}-\mu_c}{\sqrt{2\sigma_c^2}}\right)\right] \quad (18)$$

If no real intersections are found, this indicates that the particle's Gaussian distribution is entirely contained within the cell's Gaussian curve. In this case, the volume ratio $V_{p,i}$ is equal to the particle's total volume. This ensures that the entire particle volume is accurately accounted for within the respective cell, maintaining the integrity of the void fraction calculation.

$$V_{p,i} = \int_{-\infty}^{+\infty} G_p(x)\, dx = V_p \quad (19)$$

The correction factor of the particle volume ratio ($\xi_p$) is calculated as follows. In the following equation, $N_c$ is the total number of cells around particle $p$. This correction factor is applied to ensure that the sum of the volume fractions of a particle, derived from the Gaussian integrals, accurately reflects the particle's actual volume.

$$\xi_p = \frac{V_p}{\sum_{i=1}^{N_c} V_{p,i}} \quad (20)$$

The solid fraction and void fraction of cell $i$ is

$$\alpha_{s,i} = \frac{\sum_{p=1}^{N_p} \xi_p V_{p,i}}{V_{c,i}} \quad (21)$$

$$\alpha_{f,i} = 1 - \alpha_{s,i} \quad (22)$$

For calculating the void fraction at the particle location in the DEM solver, the gradient of the void fraction in the CFD solver is derived by solving the following equation for the domain at each CFD time step. The void fraction ($\alpha_{f,i}$) is defined as a scaler that is to be solved in the CFD solver.

$$\frac{\partial \alpha_{f,i}}{\partial t} = S_{\alpha_{f,i}} \quad (23)$$

$S_{\alpha_{f,i}}$ is the source term for the void fraction scalar in the CFD solver which is defined as follows.

$$S_{\alpha_{f,i}} = \frac{\alpha_{f,i}^{t} - \alpha_{f,i}^{t-\Delta t}}{\Delta t} \quad (24)$$

The gradient value for each cell is transferred to the DEM solver to estimate the void fraction at the particle's location using the following equation.

$$\alpha_{f,p} = \alpha_{f,i} + \nabla \alpha_{f,i} \cdot dx_{p,i} \quad (25)$$

Which $\alpha_{f,p}$ is the void fraction at particle $p$ location and $dx_{p,i}$ is the distance between the cell center $i$ from particle $p$.

Figure 2 illustrates the neighboring cells in the first and second layers of structured hexahedral and unstructured tetrahedral grids. These neighboring cells are crucial in determining the surrounding cells into which a particle's volume can be distributed. Each neighboring cell is evaluated for the potential distribution of the particle's volume. The first layer has six for hexahedral and four for tetrahedral cells. The second layer has eighteen for hexahedral and twelve neighbor cells for tetrahedral cell types. The number of cell neighbors in the second layer in the unstructured grid types is not definite. Note that, unlike the hexahedral or tetrahedral cells shown in Figure 1 for simplicity, the polyhedral cells used in this study may have up to 22 neighbors.

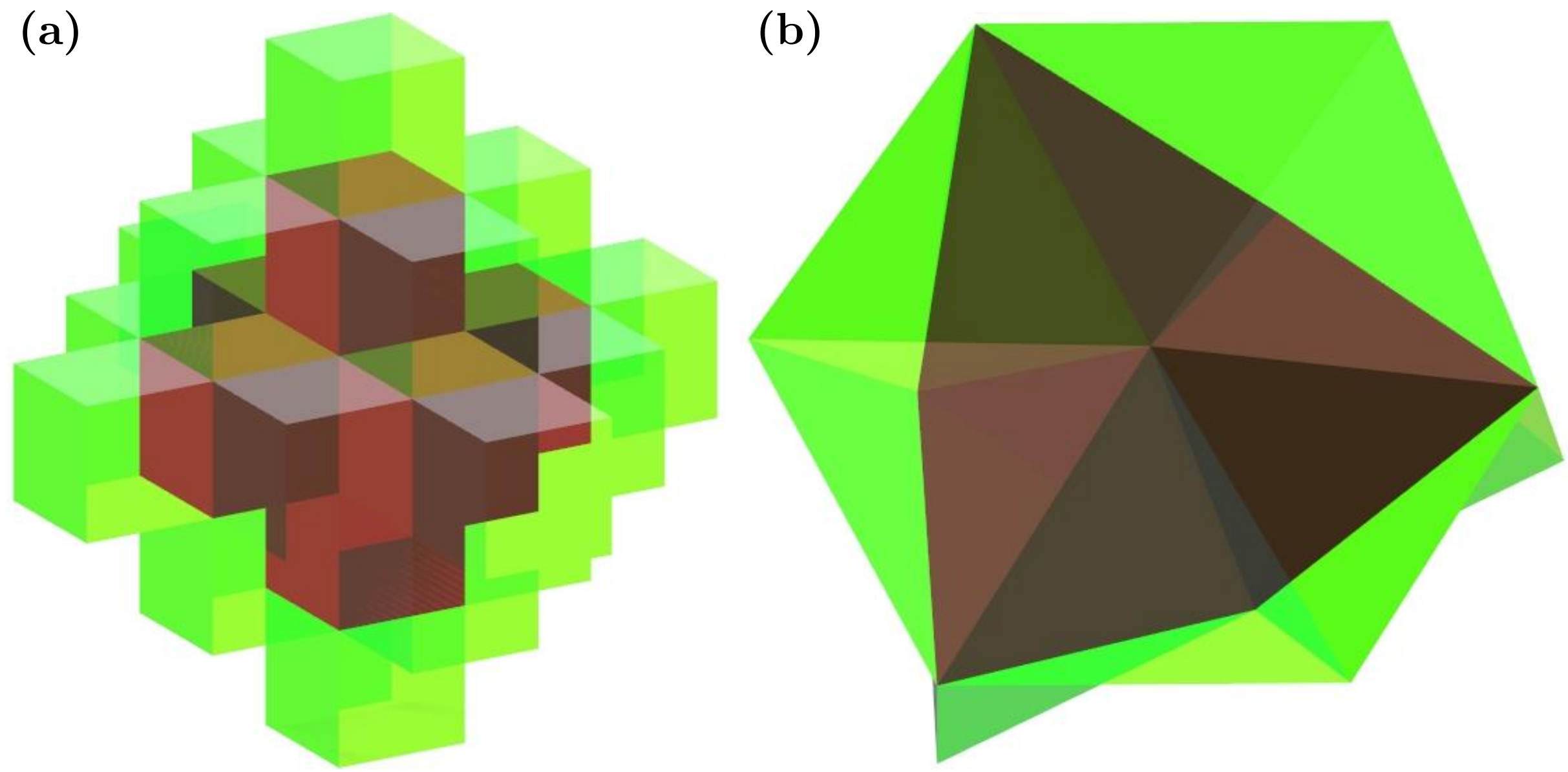

Figure 2. The neighbor cells of the first (red) and second layer (green) of hexahedral (a) and tetrahedral (b) cells. The first layer has six/four, and the second layer has eighteen/twelve neighbor cells for hexahedral/tetrahedral cell types.

The ratio of $V_{p,i}/V_p$ (volume portion of particle that belongs to a cell $i$) for the distance of $\mu_c = r_c - r_p$ (particle resides inside the spherical boundary of a cell) is depicted in Figure 3. The figure shows the ratio of $1.0$ for all relative cell volumes. The relative cell volume is the ratio of the cell volume to the particle volume ($V_c/V_p$). In the figure, the y-axis represents the Gaussian function values for a particle with a total volume $V_p = 1$. The calculations use a tuning coefficient $\phi = 0.4$, and the relative cell volumes $V_c/V_p$ are indicated in the figure. Same procedure applies to Figures 4 and 5.

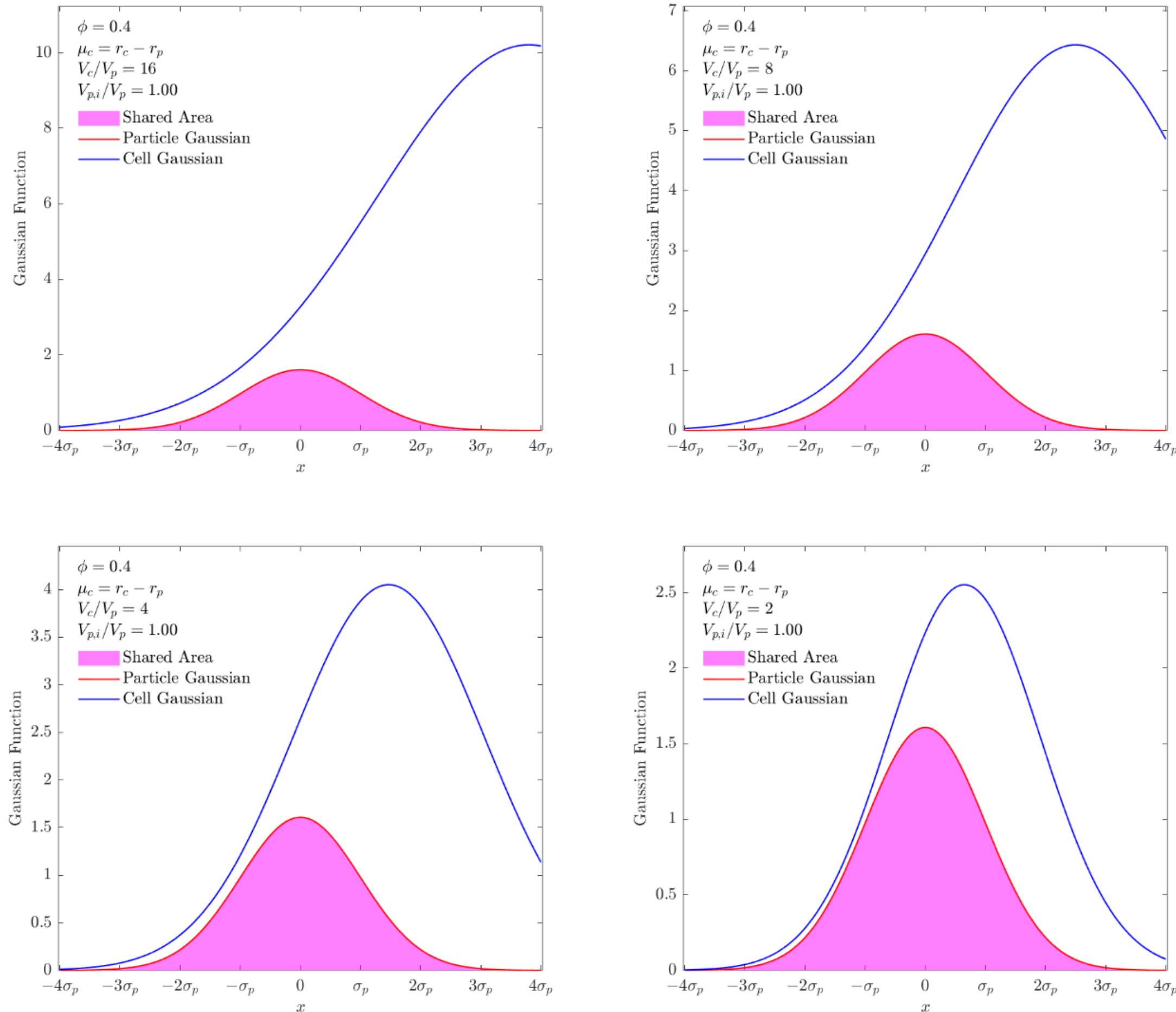

Figure 3. The ratio of $V_{p,i}/V_p$ for the distance of $\mu_c = r_c - r_p$ where particle resides inside the spherical boundary of a cell, for relative cell volumes of 2, 4, 8, and 16.

In Figure 3, where all of the particle resides inside the cell, the particle volume ratio of one ($V_{p,i}/V_p = 1$) is the objective that is achieved by GIM.

The ratio of $V_{p,i}/V_p$ for the distance of $\mu_c = r_c$ (particle resides on the radius of the spherical boundary of a cell) is shown in Figure 4. The figure shows the particle volume ratios of 0.24, 0.28, 0.34, and 0.44 for relative cell volumes of 2, 4, 8, and 16, respectively.

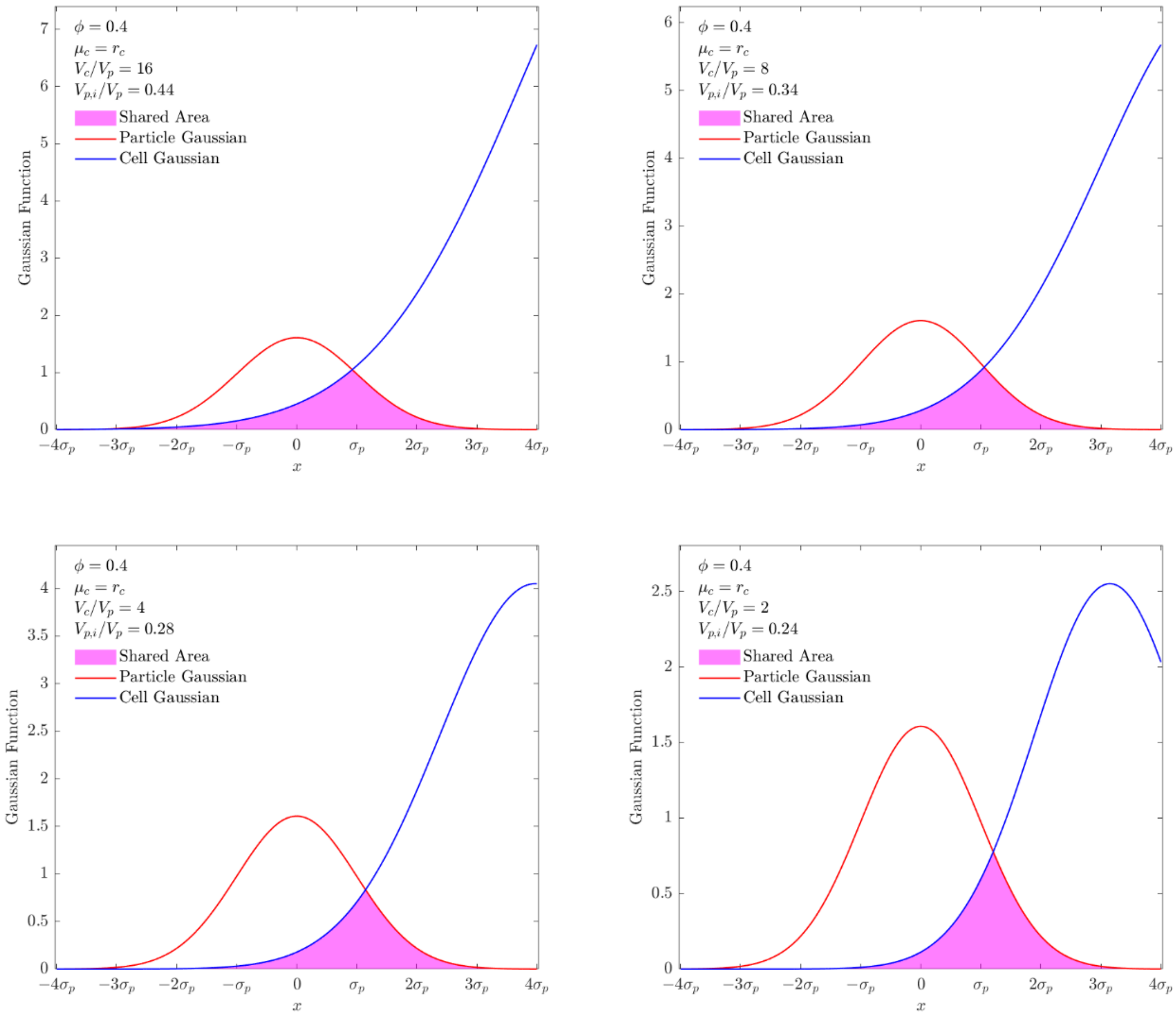

Figure 4. The ratio of $V_{p,i}/V_p$ for the distance of $\mu_c = r_c$ where particle center resides on the radius of the spherical boundary of a cell for relative cell volumes of $2$, $4$, $8$, and $16$.

In Figure 4, where the center of the particle's center resides on the cell's spatial limit (imaginary radius of the cell), the particle volume ratio of less than half ($V_{p,i}/V_p < 0.5$) is the goal. Furthermore, the ratio should increase by increasing the ratio of the volume of a cell to the volume of a particle ($V_c/V_p$). The conditions are achieved by GIM.

The ratio of $V_{p,i}/V_p$ for the distance of $\mu_c = r_c + r_p$ (particle radius touches the spherical boundary of a cell from outside) is illustrated in Figure 5. The figure shows the particle volume ratios of $0.02$, $0.03$, $0.05$, and $0.08$ for relative cell volumes of $2$, $4$, $8$, and $16$, respectively.

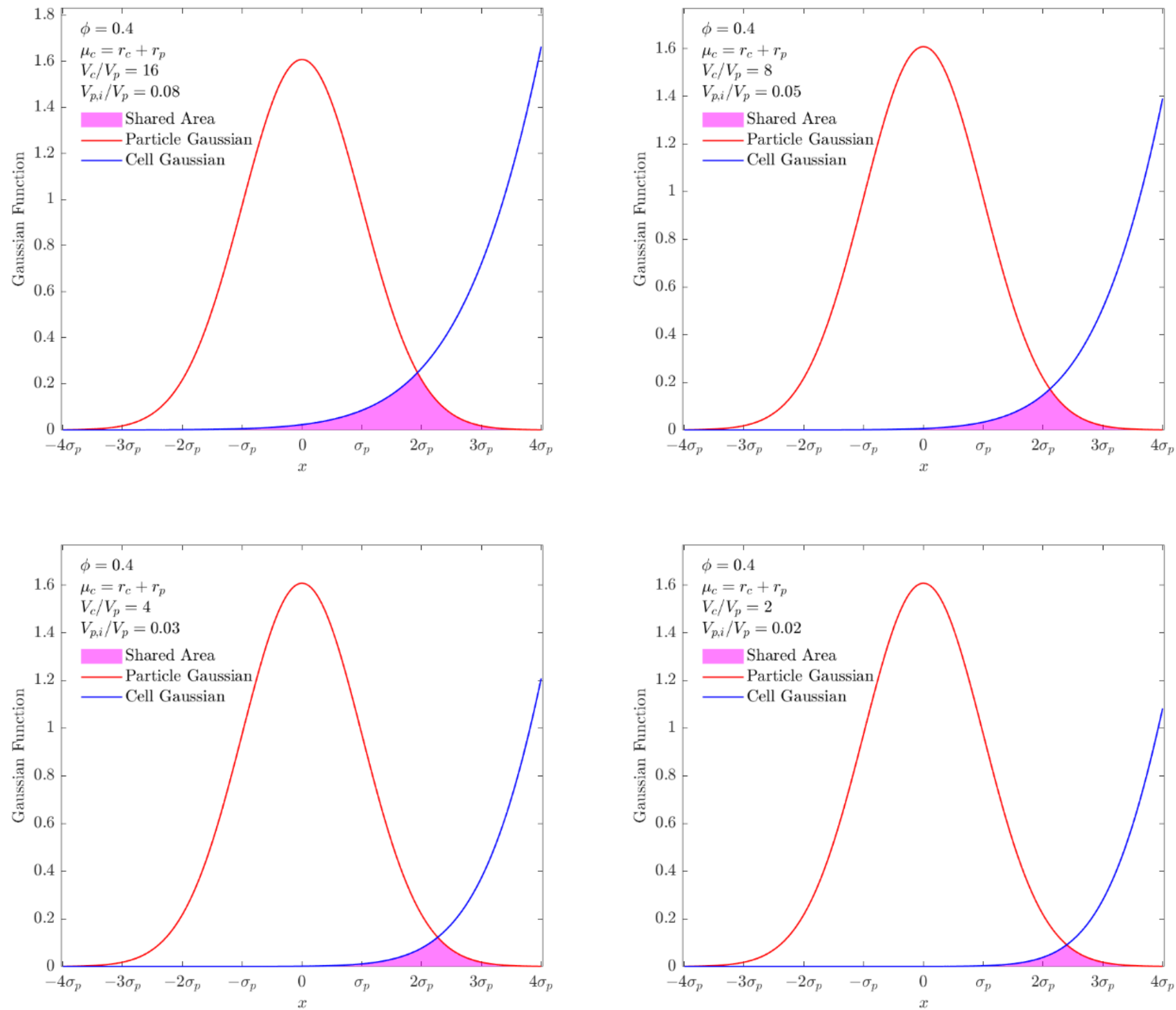

Figure 5. The ratio of $V_{p,i}/V_p$ for the distance of $\mu_c = r_c + r_p$ where particle radius touches the spherical boundary of a cell from outside for relative cell volumes of $2$, $4$, $8$, and $16$.

In Figure 5, where the entire particle volume resides outside the cell's spatial limit, the particle volume ratio approaches zero ($V_{p,i}/V_p \rightarrow 0$), which is the desired outcome. Similarly, in the previous scenario, the ratio increases as the volume ratio of the cell to the particle ($V_c/V_p$) grows. However, because the spatial boundaries defined by GIM do not perfectly align with those of a real CFD grid, particles located outside a cell have a higher probability of contributing a portion of their volume to neighboring cells as the cell volume increases.

While other methods, such as single Gaussian smoothing approaches or direct overlap calculations (often used for structured grids), often struggle with accurately distributing particle volumes in these situations—especially when dealing with irregular geometries or varying cell sizes—GIM successfully meets all the previously mentioned conditions. By employing Gaussian functions to represent both particles and cells, GIM effectively accounts for the statistical distribution of particle volumes across cells. This ensures accurate void fraction calculations across various geometries, even under conditions that typically challenge other methods.

Figure 6 illustrates the ratio of $V_{p,i}/V_p$ as a function of the relative cell volume $V_c/V_p$ for various values of the tuning coefficient $\phi$ ranging from $0.2$ to $0.8$. This figure considers three scenarios based on the particle's position relative to the cell:

- Particle inside the cell where touching cell spherical boundary ($\mu_c = r_c - r_p$)
- Particle center on the cell spherical boundary ($\mu_c = r_c$)
- Particle outside the cell ($\mu_c = r_c + r_p$)

The objective of Figure 6 is to identify a single value of $\phi$that ensures optimal performance—meaning realistic and accurate particle volume ratios $V_{p,i}/V_p$—across all three scenarios, regardless of the particle's location relative to the cell. By finding such a value of $\phi$ for varying $V_c/V_p$, The aim is to enhance the method's applicability and accuracy in different geometrical configurations.

The right column shows a detailed view of results for $\phi$ of $0.3$ to $0.5$. The $\phi$ values less than 0.36 for the situation that particle is inside the cell's spatial limits ($\mu_c = r_c - r_p$) yields particle volume ratio of less than one ($V_{p,i}/V_p < 1$) which is not realistic. For the situation where the particle's center is on the cell's spatial limits ($\mu_c = r_c$), the $\phi$s smaller than $0.3$ results in unconvincing particle volume ratios smaller than $0.1$. Also, the $\phi$s bigger than $0.5$ for the declared situation have ratios bigger than $0.5$, which is unrealistic.

Where the particle is entirely outside the cell's spatial limits ($\mu_c = r_c + r_p$), $\phi$ bigger than $0.5$ ($\phi > 0.5$) have high particle volume ratios ($V_{p,i}/V_p > 0.1$ for $V_c/V_p > 8$). The near-zero ratio is the target in this situation. Based on the plots' data, the $\phi$s between 0.36 and 0.5 ($0.36 \leq \phi < 0.5$) can calculate realistic particle volume ratios.

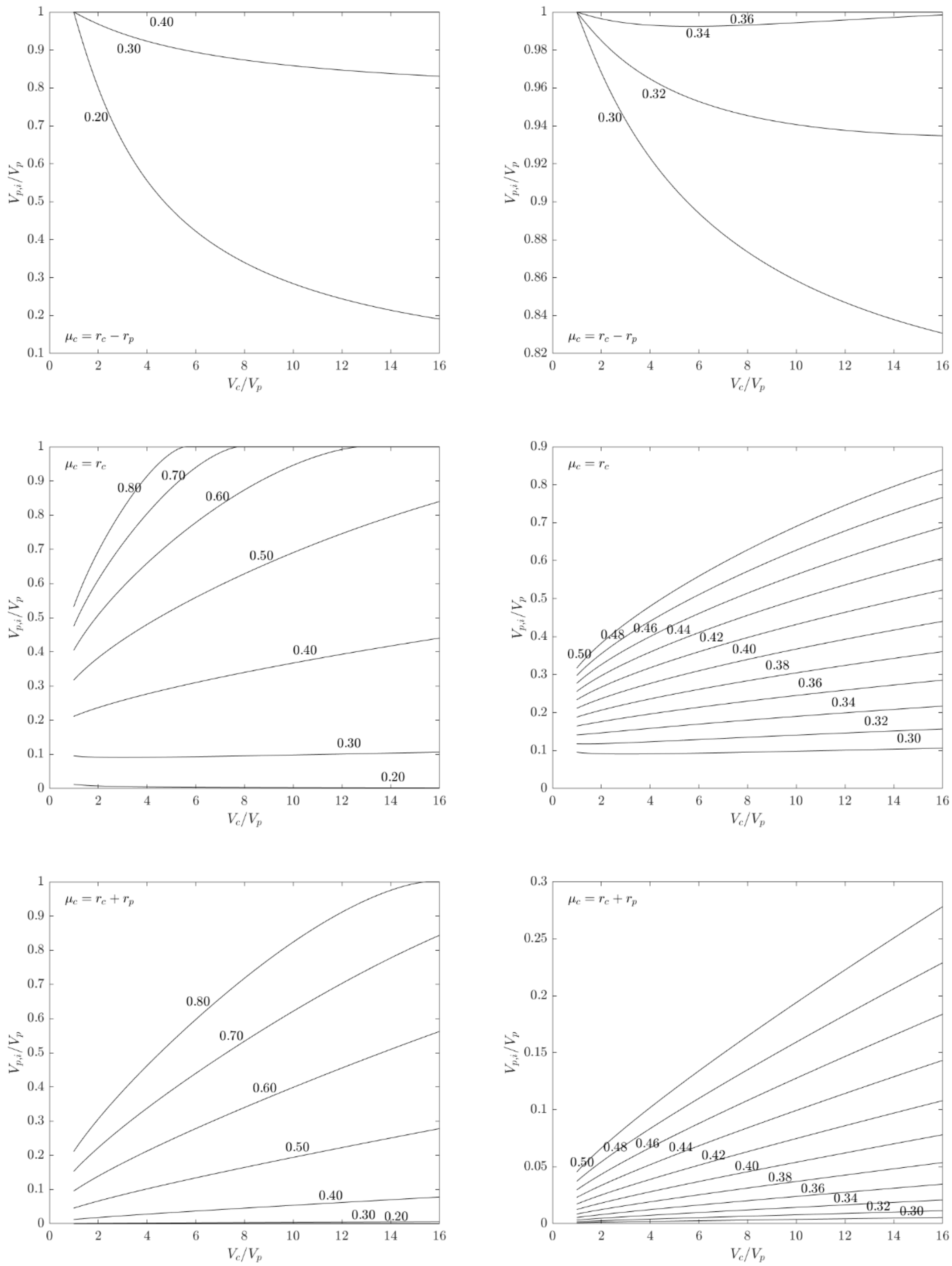

Figure 6. The ratio of $V_{p,i}/V_p$ as a function of $V_c/V_p$ for the distance of $\mu_c = r_c - r_p$, $\mu_c = r_c$, and $\mu_c = r_c + r_p$. The right column is a detailed view for $\phi$ of $0.3$ to $0.5$.

Figure 7 illustrates the ratio of $V_{p,i}/V_p$ as a function of the variable distance $\psi$, where $\mu_c = r_c + \psi r_p$, for relative cell volumes of $2$, $4$, $8$, and $16$. The right column provides a detailed view for $\phi$ values between $0.3$ and $0.5$. Values of $\phi$ less than $0.3$ are not realistic because when $\psi = -1$ (i.e., $\mu_c = r_c - r_p$), the particle is entirely inside the cell's spherical boundary, yet the calculated particle volume ratio $V_{p,i}/V_p$ is less than $1$, which is physically incorrect.

Conversely, $\phi$ values greater than $0.5$ yield particle volume ratios equal to or approaching $1$ ($V_{p,i}/V_p \to 1$) even when the particle center is located between $\mu_c = r_c - r_p$ and $\mu_c = r_c$ (i.e., $\psi \epsilon [-1,0]$), which represents the particle entirely inside the cell touching the spherical boundary or the particle partially inside the cell to the point that the particle center is at the cell's spherical boundary. This is unrealistic because, as the particle moves from entirely inside the cell towards the boundary, the volume of the particle within the cell should decrease accordingly (from $1$ to a value less than $0.5$ based on $V_c/V_p$). Therefore, the acceptable range for $\phi$ is between $0.36$ and $0.5$, where the particle volume ratio behaves realistically. Any minor inaccuracies within this $\phi$ range are negligible and can be corrected using the following tuning method and Equation (21).

For determining the optimum tuning coefficient $\phi$ for a particle-cell pair, it can be assumed that a particle is only cut by a plane, and the sliced volume is called a cap. The analytical ratio for the volume of the cap to the volume of a particle is defined as follows.

$$\frac{V_{p,i}}{V_p} = \frac{\pi {r_p}^3 (1+\psi)^2 \left(1 - \frac{1+\psi}{3}\right)}{\frac{4}{3} \pi {r_p}^3} \tag{26}$$

The analytical ratio of a particle's cap volume to particle volume ($V_{p,i}/V_p$) is illustrated by the red line in Figure 7.

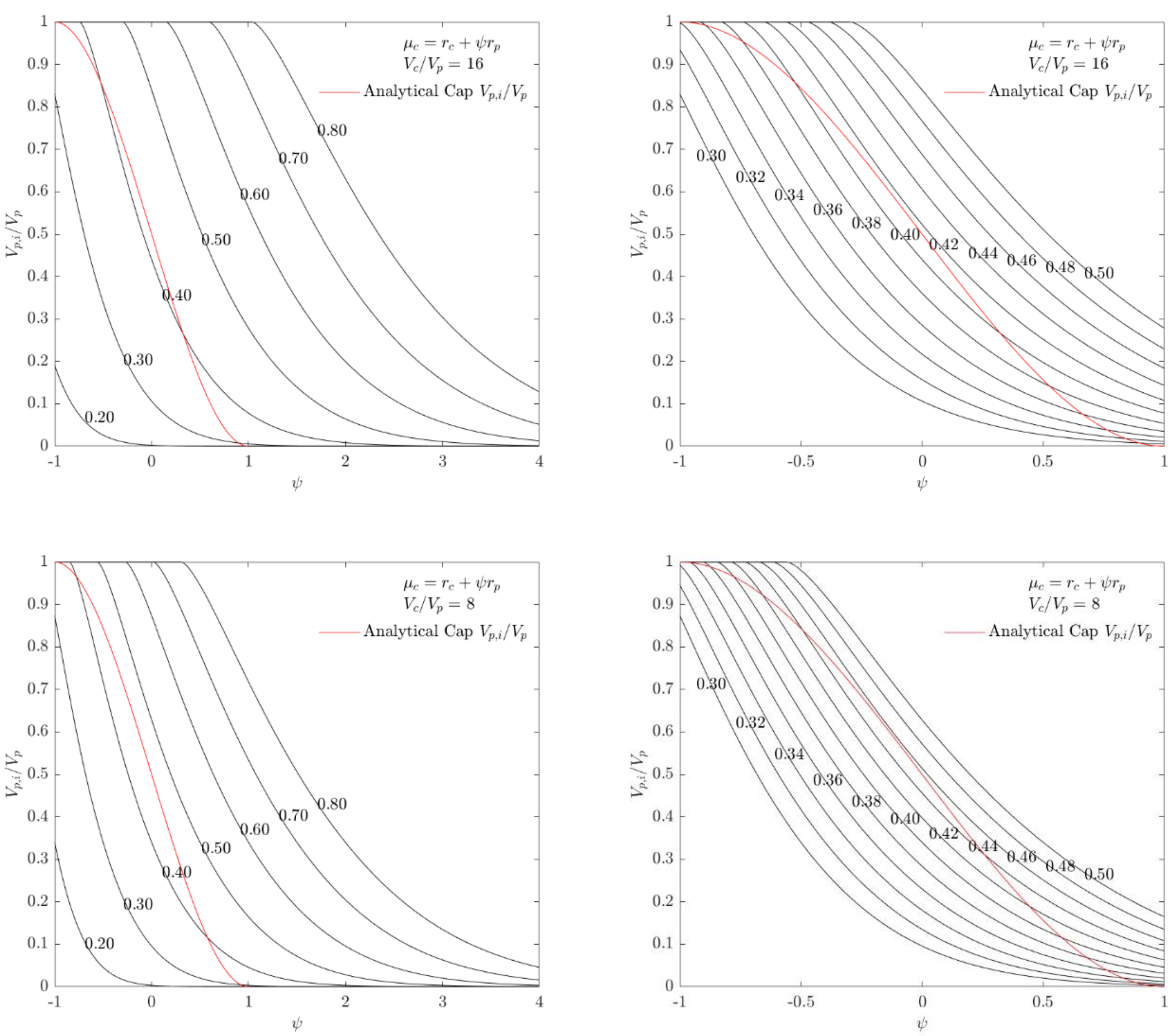

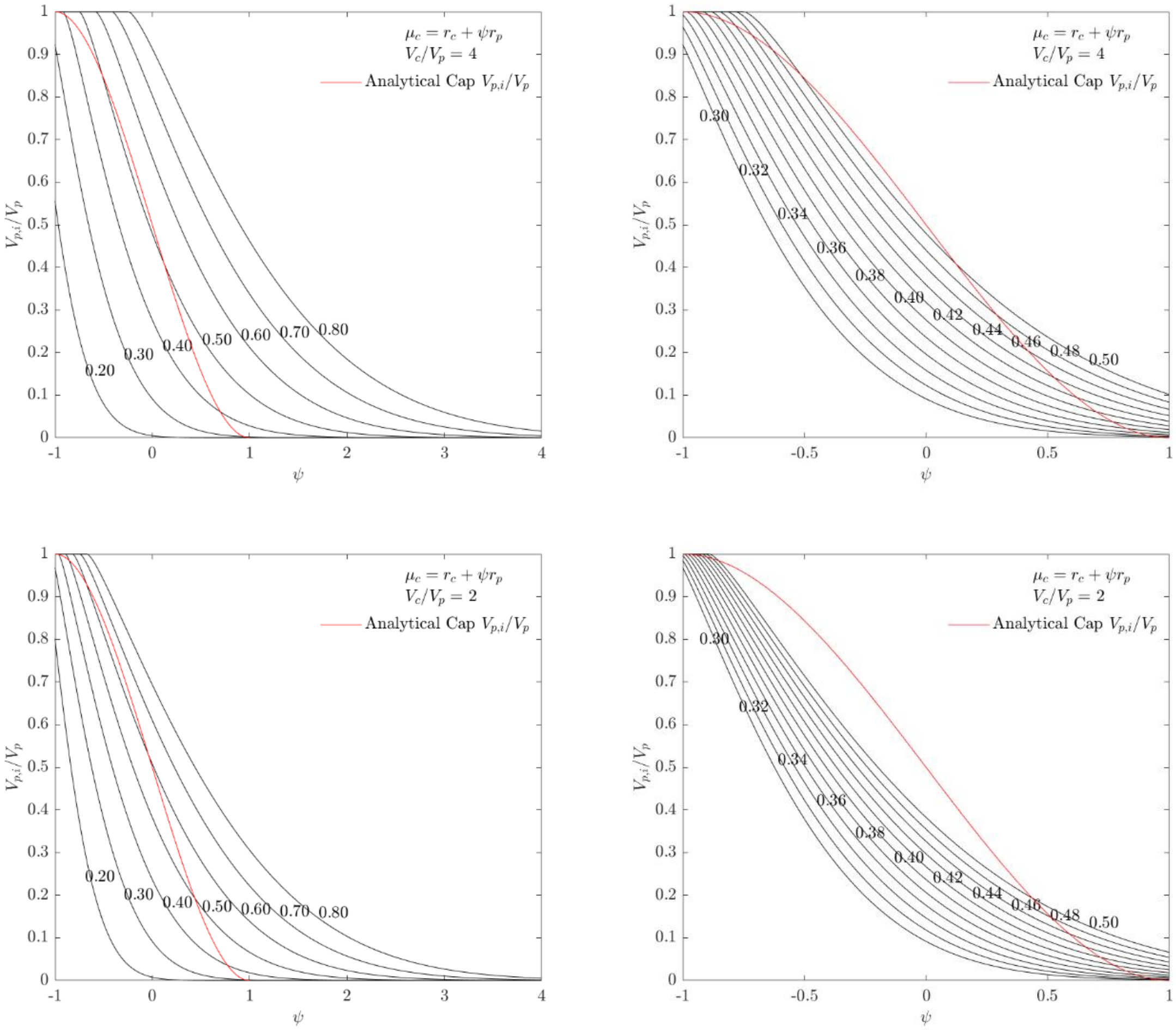

Figure 7. The ratio of $V_{p,i}/V_p$ as a function of variable distance $\psi$ coefficient ($\mu_c = r_c + \psi r_p$) for relative cell volumes of 2, 4, 8, and 16. The right column is the zoomed-in values for the $\phi$ of 0.3 to 0.5.

For each $V_c/V_p$, the optimum $\phi$ is determined by the minimum mean squared error of the GIM curves and the analytical ratio curve. The GIM curve with the optimum $\phi$ is considered the best fit for the analytical ratio curve. The optimum $\phi$s as a function of $V_c/V_p$ and the fitted trendline is illustrated in Figure 8. The trendline equation is defined as follows.

$$\phi = 0.579\left(\frac{V_p}{V_c}\right)^{0.132} \tag{27}$$

The optimum $\phi$ correlation is used for each particle-cell pair. This can make the method independent of changes in cell sizes in the CFD grid.

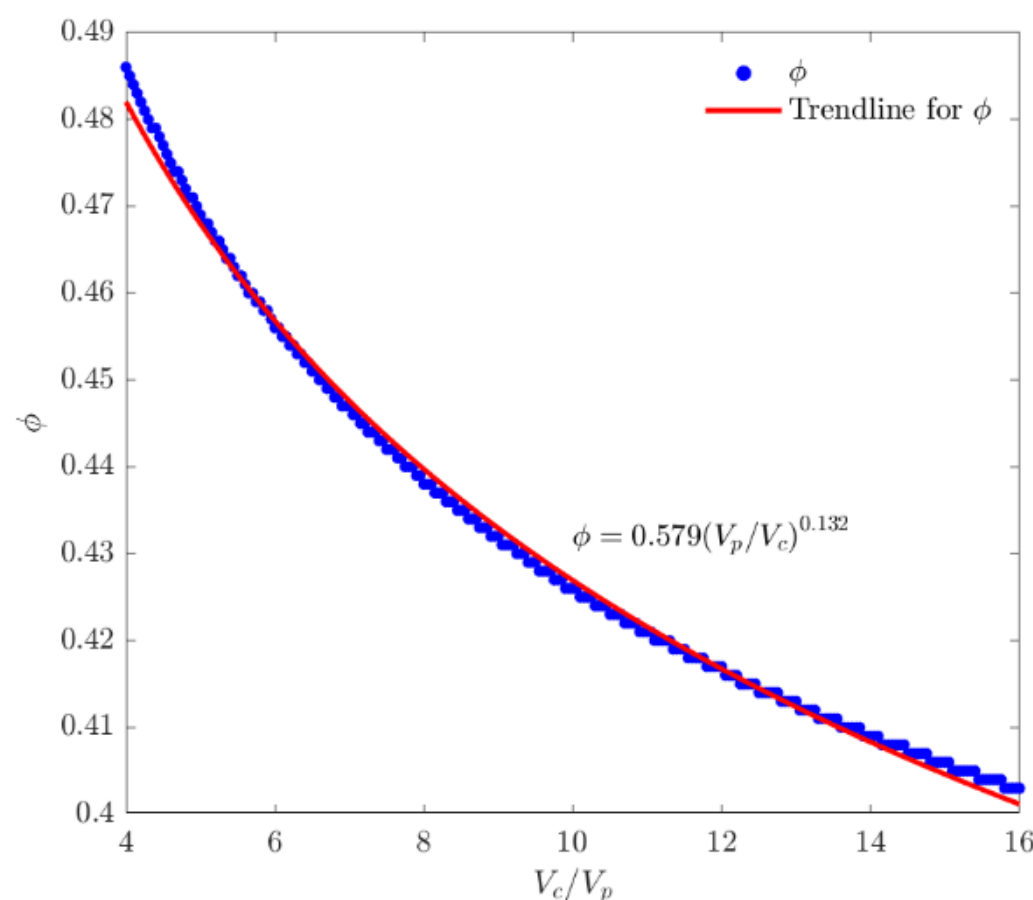

Figure 8. Optimum $\phi$s as a function of $V_c/V_p$ and the trendline.

# Results

The experiment conducted by Wachem et al. (2001), which validated their Lagrangian–Eulerian simulations of the fluidized bed, is used to validate the GIM for calculating void fraction [34]. Parameters of Wachem et al. fluidized bed, properties of particles, and fluid are presented in Table 1. The snapshots of the fluidized bed, pressure fluctuations, power spectral density (PSD) as a function of the pressure fluctuation frequency, void fraction fluctuations, and bed height fluctuations from the referenced experiment are compared with the simulation results of the current study.

Table 1. Parameters of Wachem et al. (2001) fluidized bed, properties of particles and fluid

| Simulation parameters | Notation | Value |
|---|---|---|
| **Bed** | | |
| Width ($m$) | $W$ | 0.09 |
| Transverse thickness ($m$) | $T$ | 0.008 |
| Height ($m$) | $H$ | 0.5 |
| **Particles** | | |
| Total mass ($kg$) | $\sum m_p$ | 0.039 |
| Diameter ($mm$) | $d_p$ | 1.545 |
| Density ($kg/m^3$) | $\rho_s$ | 1150 |
| Young's modulus ($Pa$) | $Y$ | $1.2 \times 10^5$ |
| Poisson's ratio | $v$ | 0.33 |
| Coefficient of normal restitution | $e$ | 0.9 |
| Coefficient of sliding friction | $\mu_s$ | 0.3 |
| **Fluid (air)** | | |
| Gas density ($kg/m^3$) | $\rho_f$ | 1.28 |
| Gas inlet superficial velocity ($m/s$) | $\overline{\boldsymbol{v}}_f$ | 0.9 |
| Gas viscosity ($Pa.s$) | $\mu_f$ | $1.7 \times 10^{-5}$ |

Simulations were conducted using two structured hexahedral grids (Grids A and B) and three unstructured polyhedral grids (Grids C, D, and E) with varying cell sizes. The details of these grids are provided in Table 2. To enable direct comparison, two of the unstructured grids were designed with cell sizes approximately equal to those of the structured grids. Specifically, Grid D (unstructured) has a cell size similar to Grid A (structured), and Grid E (unstructured) corresponds in size to Grid B (structured). Additionally, a finer unstructured grid, Grid C, was included to investigate the effectiveness of the GIM at higher resolutions.

Table 2. Grid information of simulation cases for Wachem et al. (2001) experiment

| Grid Name | Grid Type | $N_w \times N_T \times N_H$ | $N_{cells}$ | $\sqrt[3]{V_{cell}}/d_p$ |
|---|---|---|---|---|
| Grid A | structured | $33 \times 3 \times 181$ | $17919$ | $1.76$ |
| Grid B | structured | $26 \times 2 \times 140$ | $7280$ | $2.38$ |
| Grid C | Polyhedral | $-$ | $25605$ | $1.56$ |
| Grid D | Polyhedral | $-$ | $17840$ | $1.76$ |
| Grid E | Polyhedral | $-$ | $7302$ | $2.37$ |

The simulations were conducted over a total duration of $10$ seconds. To ensure reliable results, the first $2$ seconds, representing the initial stabilization phase of the fluidized bed is excluded from the analysis. The snapshots of the simulation cases of the fluidized bed and Wachem et al.'s experiment are illustrated in Figure 9.

**A** (Structured)

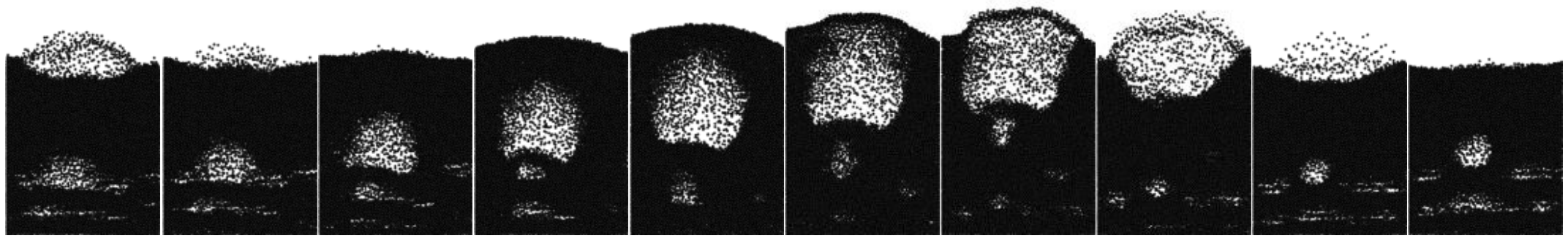

**B** (Structured)

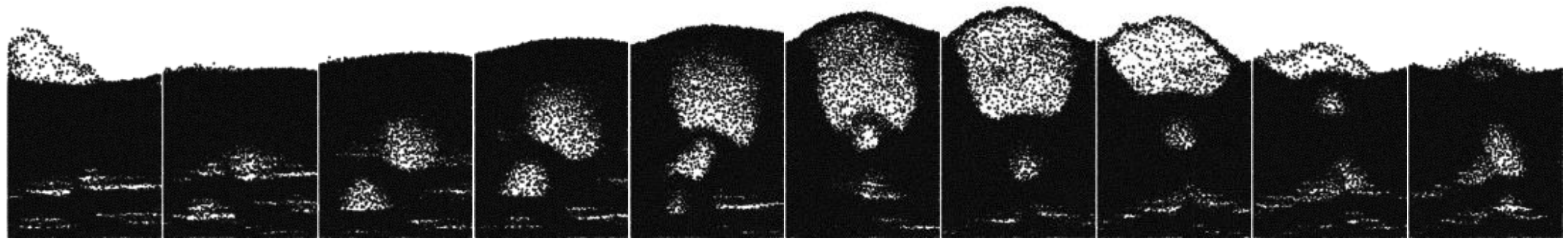

**C** (Unstructured)

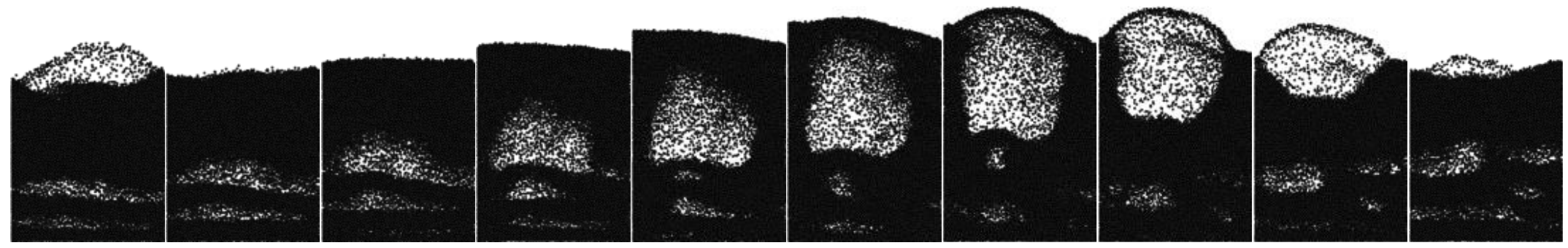

**D** (Unstructured)

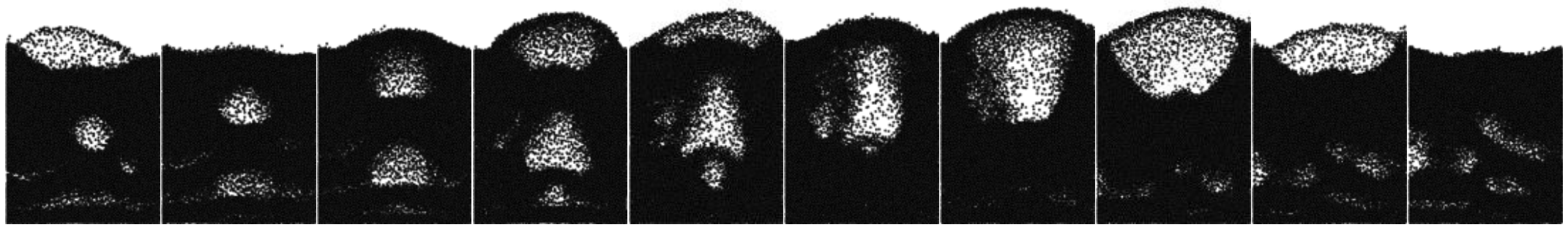

**E** (Unstructured)

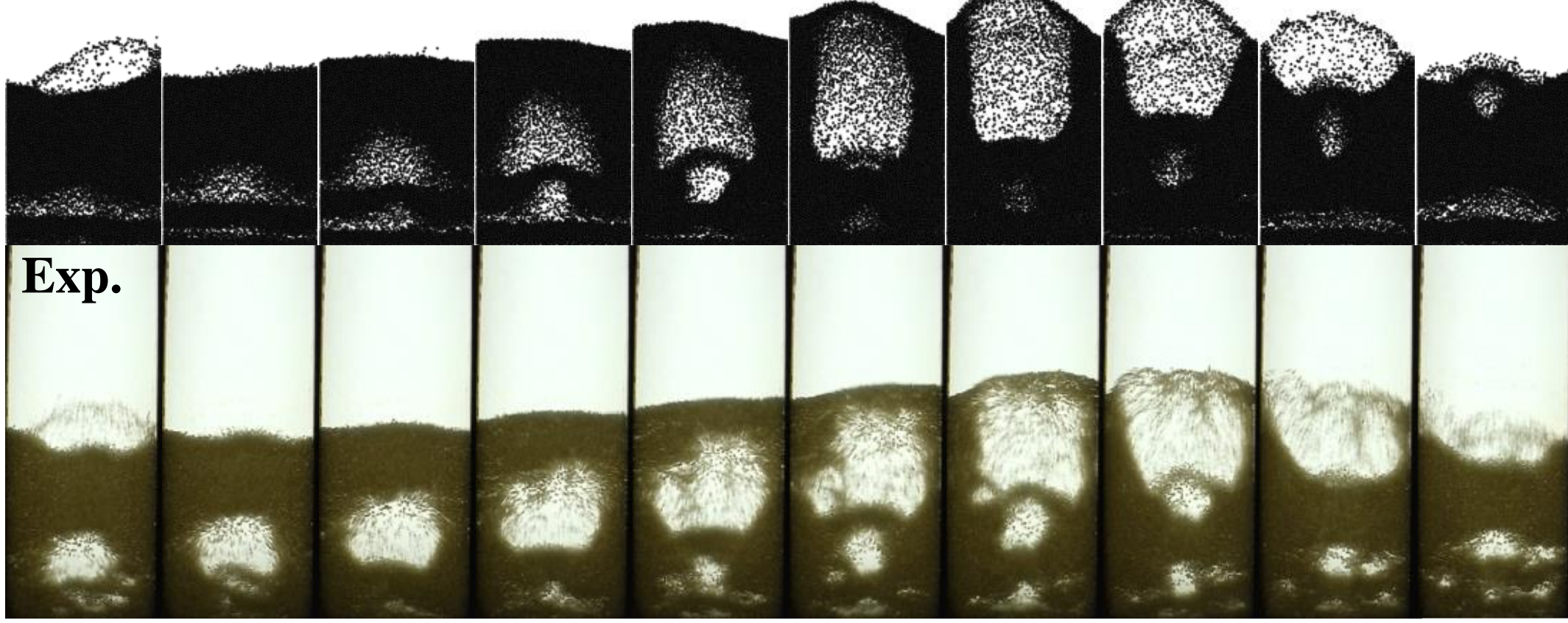

$t_0$ $t_0$+0.04s $t_0$+0.08s $t_0$+0.12s $t_0$+0.16s $t_0$+0.20s $t_0$+0.24s $t_0$+0.28s $t_0$+0.32s $t_0$+0.36s

Figure 9. Snapshots of the simulation cases of the fluidized bed and Wachem et al.'s experiment with the superficial velocity of $0.9\ m/s$.

As illustrated in Figure 9, all the structured and unstructured grid simulations resemble the experimental snapshots. At time $t_0 + 0.32\ s$, the bubble shapes in the polyhedral unstructured grid simulations (Grids C, D, and E) are notably similar to those observed in the experiment. Specifically, while Grids A and D have the same resolution (cell size approximately 1.76 times the particle diameter), Grid D (polyhedral) does not show a significant advantage over Grid A (structured) at this higher resolution. This suggests that at finer resolutions, the benefits of using polyhedral meshes are not particularly pronounced.

However, at lower resolutions, differences become more apparent. Grid E (polyhedral), which has a coarser cell size similar to Grid B (structured) at approximately 2.37 times the particle diameter, exhibits bubble formations that more closely resemble the experimental observations compared to Grid B. This indicates that the polyhedral mesh in Grid E captures the bubble dynamics more effectively at this resolution.

Moreover, even with the highest resolution in the study—Grid C (polyhedral) with a cell size of approximately 1.56 times the particle diameter—the simulation maintains realistic bubble formations similar to the experiment. This demonstrates that GIM can be effectively applied to higher-resolution grids without overly smoothing the data or losing realistic behavior. The ability of GIM to handle fine meshes while preserving accurate particle-fluid interactions is a significant advantage, ensuring that simulations remain true to physical phenomena across varying resolutions.

The Contour of void fraction ($\alpha_f$) and its partial derivatives to $x$ axis ($\partial \alpha_f / \partial x$) and $z$ axis ($\partial \alpha_f / \partial z$) is shown in Figure 10. The gradient of the void fraction is calculated by Equation (23) in the CFD solver. The contours are related to time $t_0 + 0.20\ s$ in Figure 9. As mentioned earlier, the gradient of the void fraction is transferred to the DEM solver for precise and efficient void fraction estimation at the particle's location.

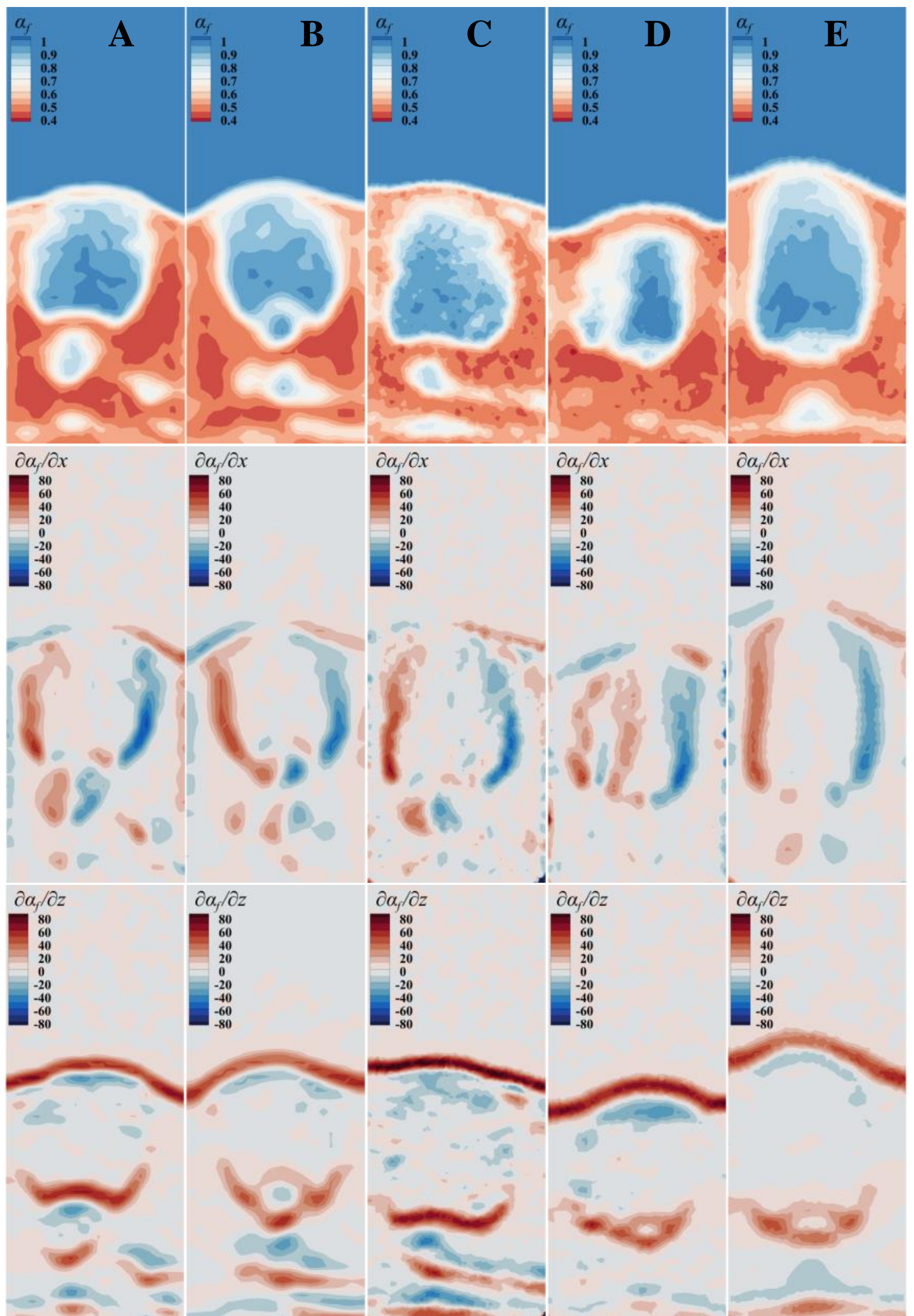

Figure 10. Contour of void fraction ($\alpha_f$) and its partial derivatives to $x$ axis ($\partial\alpha_f/\partial x$) and $z$ axis ($\partial\alpha_f/\partial z$). The gradient of the void fraction is calculated by Equation (23) in the CFD solver. The contours are related to time $t_0 + 0.20\ s$ in Figure 9 in the center plane.

Figure 10 shows the contours of void fraction ($\alpha_f$) and its partial derivatives with respect to the x-axis ($\partial\alpha_f/\partial x$) and z-axis ($\partial\alpha_f/\partial z$), calculated using Equation (23) in the CFD solver at time $t_0 + 0.20\ s$ (corresponding to Figure 9). As previously mentioned, the gradient of the void fraction is transferred to the DEM solver for precise and efficient estimation of the void fraction at the particle's location, which is critical for calculating drag forces.

As illustrated in Figure 10, the boundaries of the bubbles are clearly delineated by the partial derivatives of the void fraction. When comparing the different meshes, we observe that the structured grids exhibit more homogeneous void fraction contours in the center plane. In contrast, the unstructured polyhedral grids, particularly at higher resolutions (Grids C and D), display sharper gradients and slightly more noise. This is expected because higher-resolution grids capture the presence of particles within cells more precisely, leading to more abrupt changes in void fraction from one cell to another.

Despite these differences, all meshes demonstrate appropriate behavior, and no unrealistic void fractions are observed in the contours. Importantly, there is no accumulation of solid fraction at the boundaries; the boundary regions exhibit lower solid fractions, which aligns with physical expectations in real-world scenarios. This indicates that GIM effectively handles boundary conditions and maintains realistic void fraction distributions across different grid types and resolutions.

Figure 11 illustrates a subset of the pressure fluctuation time series recorded at $45\ mm$ above the distributor. In the structured case A, it is evident that the pressure fluctuations frequency is higher than those observed in the experiment. However, the amplitude of the pressure fluctuation is approximately in the range of the experiment. This suggests that the bubbles in the simulation are almost the same size as in the experiment. The relative pressure peaks for the structured case B, which uses a coarser grid, are less prominent and more irregular. However, the amplitude of the fluctuations is closer to the experimental values. This indicates that the voids in this simulation are more similar to those perceived in the experiment. The polyhedral cases C and D have similar behavior for the relative pressure fluctuations. The amplitude of cases C and D are relatively stronger than that of case B and the experimental data. The time scale of the pressure fluctuations derived from the polyhedral cases moderately aligns with the experiment. However, the amplitude of case E is noticeably higher than that of the experiment, which means that larger bubbles in the bed are generated than in the experimental observation. Furthermore, the pressure fluctuation frequency is lower than in other structured and unstructured cases and is very close to the experiment's frequency. A comprehensive comparison of frequency and the amplitude of the pressure fluctuation to the experiment's data is presented in the PSD plot (Figure 12).

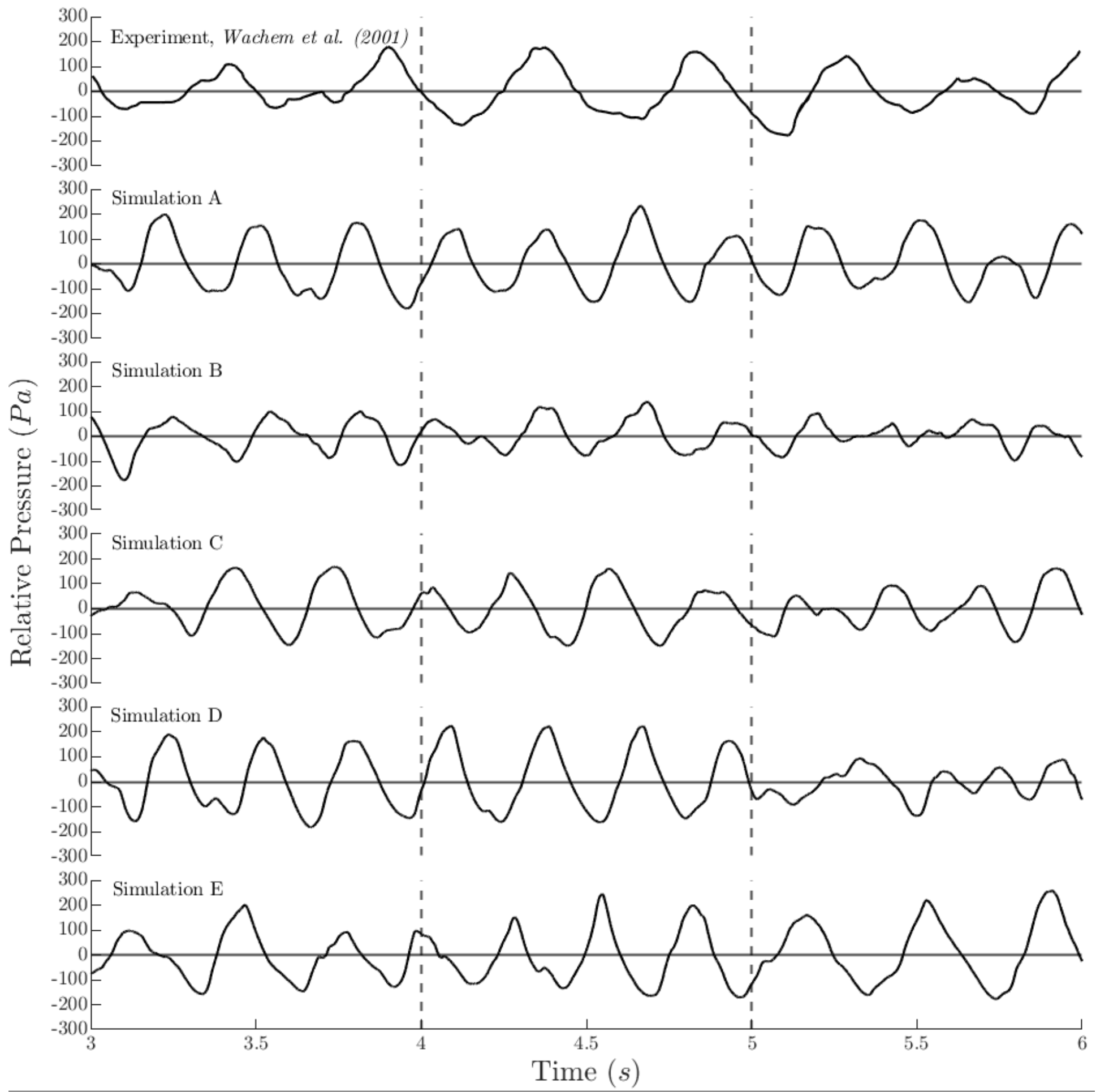

Figure 11. Relative pressure fluctuation at the height of 45 mm, comparing five different simulation strategies with the experimental data from Wachem et al.

The PSD of the entire pressure fluctuation time series, measured at $45\,mm$ above the distributor, is presented in Figure 12, using the same calculation procedure as Wachem et al. The pressure fluctuation is recorded every $1.0E-4\,s$ or in $1.0E4\,Hz$ of frequency. The sampling is downscaled to $1.0E3\,Hz$. In the referenced experiment a filter of 314 Hz was applied to the pressure fluctuations; however, the data from the current study are raw with no filtration.

It can be observed in Figure 12 (A) that the dominant frequency for the experimental data is around $2.5\,Hz$. The dominant frequency is induced by the bubble behavior. The dominant frequency for all cases except case E is approximately $3.5\,Hz$. However, the dominant frequency for case E is about $3\,Hz$ which is the closest to the experiment's data. On the other hand, all cases overestimate the relative pressure amplitude at the dominant frequency. Among all cases, the structured grid B (the coarser structured grid) has a more similar relative pressure amplitude of the dominant frequency to the experiment. The unstructured case E has a relative pressure amplitude at the dominant frequency that is higher than all cases. The coarser grids of E and B result in more noisy PSD plots in frequencies higher than $3\ Hz$. The PSD plots for all cases overlap each other and resemble the experiment's data, except for the frequencies lower than $3\ Hz$ for case E. This indicates that the GIM is grid-independent and produces similar results for both structured and unstructured grids.

Figure 12 (B) is the PSD plots of the cases without employing the optimization method. By comparing Figure 12 (A) and (B), it is evident that the optimization method helps with the grid-independency of GIM. Furthermore, the simulations without the optimization method have noisy fluctuation and less dominant frequencies, which leads to irregular and deformed bubbles. The simulations without the optimization method are sensitive to the grid resolution and type.

The decline in the PSD curve in Figure 12 for all cases at frequencies higher than $5\ Hz$, which follows a power law drop, is characteristic of gas-solid fluidized beds. The concordance between experimental and simulation data is accurate in this matter [34, 35].

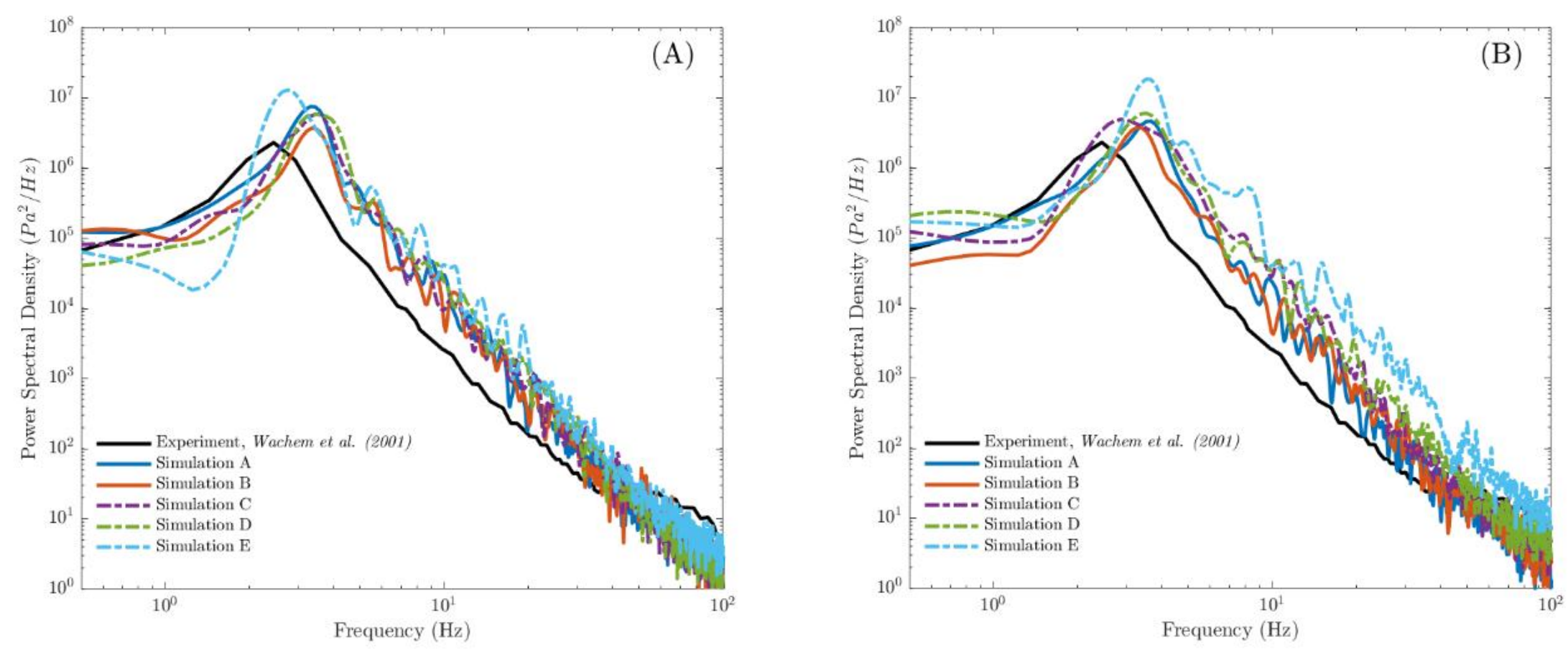

Figure 12. The PSD of the relative pressure fluctuations as a function of frequency for the cases with optimization (A) and without optimization (B) and Wachem et al.'s experiment is measured at a height of 45 mm. The structured grids are shown with solid lines, and the unstructured grids are shown with dash-dotted lines in different colors.

Figure 13 illustrates the void fraction fluctuations, averaged over time in a horizontal plane at $45\ mm$ above the distributor, for the simulation cases compared with the experimental data. It should be noted that the referenced experiment calculated the void fraction in two dimensions by measuring light intensity derived from video recording. The transition from two dimensions to three-dimensional void fraction is done by Hoomans et al.'s correlation presented in the referenced study [8, 34]. The correlation can introduce inaccuracies for the measured void fraction in three dimensions.

Cases A and C, with the finer grid compared to other structured and unstructured grids, have broader peaks compared to other cases, and they are more similar to the experiment's data. This can be due to the higher number of cells at $45\,mm$ for averaging the void fraction. Consequently, the coarser grids have noisy fluctuations and sharper peaks. The major peaks in the void fraction fluctuations (by neglecting the small noises) for the time duration of $(3, 4)\ s$ match the peaks of the relative pressure fluctuations in Figure 11, which contains three dominant peaks. This is more obvious in case A, which relates to the time the fluid phase forms and lifts a bubble.

The amplitude of the fluctuations is associated with the width of the bubbles. The standard deviation of the void fraction plot is presented in Table 3. Case E has the highest void fraction amplitude, meaning larger bubbles than the others. The relative pressure fluctuation plot (Figure 11) and the PSD plot (Figure 12) indicate larger bubbles of case E by having higher relative pressure amplitude at the dominant frequency.

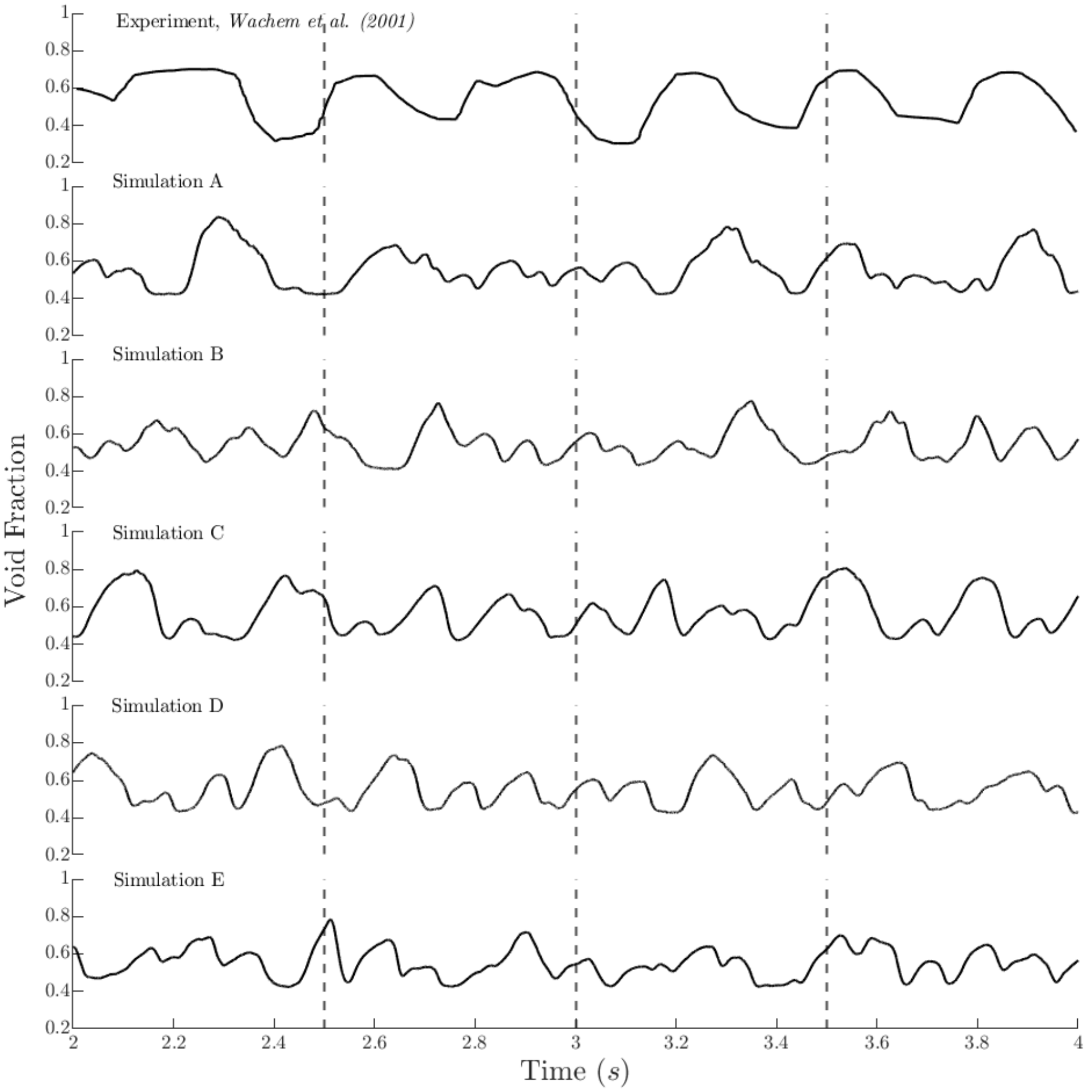

Figure 13. Void fraction fluctuation at the height of 45 mm, comparing five different grid strategies with the experimental data from Wachem et al.

Figure 14 depicts the bed height over time resulting from the simulation cases and the experiment. The shape of the bed height fluctuations is similar to the experimental data, especially for unstructured polyhedral grids. The structured grids of A and B have sharp peaks; however, the unstructured polyhedral grids (cases C, D, and E) have round peaks similar to the experiment's data. The polyhedral grids have a better resemblance due to the more accurate fluid field gradients since they have more neighbor cells than the structured grids. Table 3 presents the mean and standard deviation of the bed height. Case E (the same resolution as case B) has the closest mean and standard deviation of bead height fluctuation to the experimental data. Likewise, the shape of the fluctuations in bed height in case D more closely resembles the experimental data than in case A despite having the same resolution. Thus, polyhedral grids can produce more realistic results than structured grids with the same resolution.

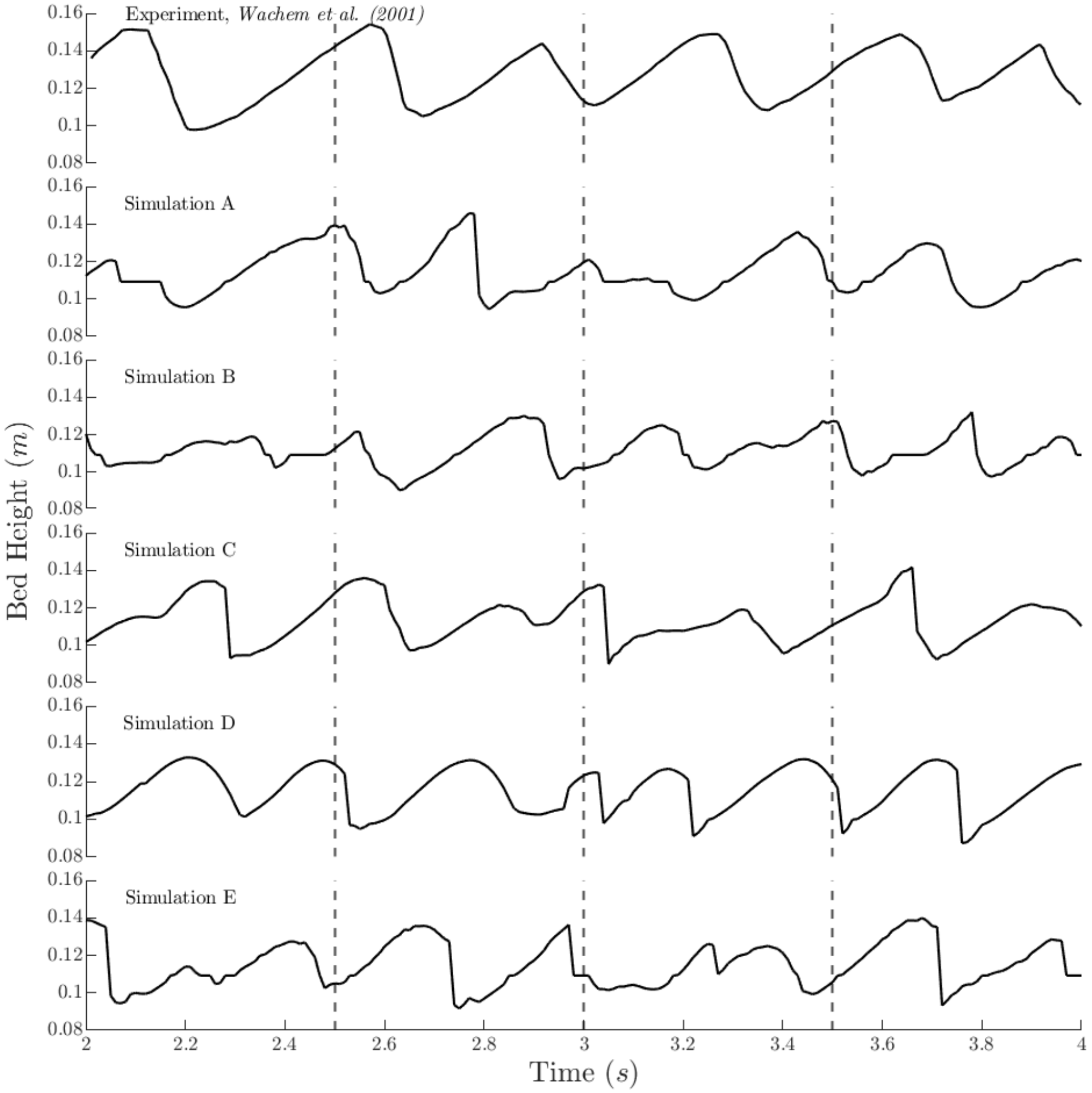

Figure 14. Bed height fluctuation at the height of 45 mm, comparing five different simulation strategies with the experimental data from Wachem et al.

The bed height's mean and standard deviation for Wachem et al.'s experiment and the simulations are presented in Table 3. As mentioned earlier, the higher standard deviation of the void fraction in case E indicates larger bubbles, which can also be seen in the higher standard deviation of the bed height. Furthermore, the mean and standard deviation of the bed height in case E are closer to the experimental results. In contrast, the structured grid of case B, with the same resolution as case E, has the mean and standard deviation farthest from the experimental results. This shows further evidence that using a polyhedral grid with the same resolution as structured grids can produce more realistic simulation results. Similarly, the unstructured polyhedral grid D has a mean bed height closer to the experimental data than the structured grid A at the same resolution. However, the standard deviation of the bed height for grid D is slightly less favorable.

Table 3. The mean and the standard deviation for Void fraction and bed height for Wachem et al.'s experiment and the simulation cases.

| | Void fraction $[-]$ | Bed height $[\boldsymbol{m}]$ |
|---|---|---|
| Experiment Wachem et al. | Not provided | $0.12 \pm 0.019$ |
| Simulation A | $0.550 \pm 0.092$ | $0.1133 \pm 0.0142$ |
| Simulation B | $0.552 \pm 0.069$ | $0.1119 \pm 0.0119$ |
| Simulation C | $0.555 \pm 0.095$ | $0.1141 \pm 0.0126$ |
| Simulation D | $0.549 \pm 0.082$ | $0.1156 \pm 0.0121$ |
| Simulation E | $0.555 \pm 0.102$ | $0.1160 \pm 0.0161$ |

## Conclusion

In this study, the Gaussian Integral Method (GIM) was introduced as a novel approach for calculating void fractions in CFD-DEM simulations of particulate media. The method addresses the challenges of accurately determining void fractions, especially in simulations involving complex geometries and various grid types. The key findings are summarized as follows:

- Versatility Across Grid Types: GIM was demonstrated to be a versatile method applicable to various grids, including structured and unstructured polyhedral meshes, without requiring special boundary treatments as other Gaussian methods do. This enhances its applicability to complex geometries common in industrial simulations.
- Grid Independence Through Optimization: An optimization technique was developed to make GIM independent of grid resolution and type. By adjusting the standard deviation of the Gaussian functions based on particle and cell volumes, GIM adapts to variations in grid resolution and particle sizes, ensuring consistent and accurate void fraction calculations.
- Validation Against Experimental Data: The method was validated against experimental data from Wachem et al., showing that employing GIM in CFD-DEM simulations produces realistic results for fluidized bed dynamics. Accurate representations of relative pressure fluctuations, bubble formation, and bed height variations were achieved, closely resembling experimental observations.
- Advantages of Polyhedral Grids: Polyhedral grids using GIM demonstrated advantages over structured grids of equivalent resolution, producing results that more closely match experimental data. This improvement is attributed to the polyhedral grids' higher number of neighboring cells, leading to more accurate calculations of fluid field gradients.

- Accurate Void Fraction Gradients: GIM enabled precise calculation of void fraction gradients, which is crucial for accurately estimating local void fractions at particle locations in the DEM solver. This accuracy enhances the modeling of particle-fluid interactions and contributes to more reliable simulation results.

In conclusion, the Gaussian Integral Method provides an effective and robust solution to the complex problem of calculating void fractions in CFD-DEM simulations, particularly for simulations involving complicated geometries with polyhedral meshes. Its versatility, accuracy, and ability to operate without special boundary treatments make it a valuable tool for enhancing the fidelity of multiphase flow simulations in various industrial applications. Future research could focus on further optimizing the method by employing distinct standard deviations for particles and cells or by extending its application to more complex multiphase systems.

## Nomenclature

| | |
|---|---|
| $d_p$ | Diameter of particle |
| $e$ | Coefficient of normal restitution |
| $\mathcal{F}$ | Function symbol |
| $\boldsymbol{F}_n$ | Normal force vector |
| $\boldsymbol{F}_t$ | Tangential force vector |
| $\boldsymbol{F}_d$ | Drag force vector |
| $\boldsymbol{g}$ | Gravity vector |
| $I$ | Moment of inertia |
| $\mathbf{I}$ | Identity matrix |
| $m$ | Mass |
| $P$ | Pressure |
| $\boldsymbol{r}$ | Particle’s center of mass to the collision contact point vector |
| $Re_d$ | Reynolds number for drag |
| $\boldsymbol{S}_p$ | Momentum source term vector |
| $U_f$ | Superficial fluid velocity |
| $V_p$ | Volume of particle |
| $\boldsymbol{v}_f$ | Fluid velocity vector |
| $Y$ | Young's modulus |

*Greek symbols*

| | |
|---|---|
| $\alpha_f$ | Void fraction |
| $\alpha_s$ | Solid fraction |
| $\beta_{BVK}$ | BVK inter-phase momentum exchange coefficient |
| $\mu_f$ | Fluid shear viscosity |
| $\rho_f$ | Fluid density |
| $\rho_s$ | Solid density |
| $\boldsymbol{\tau}_f$ | Fluid viscous stress tensor |
| $\boldsymbol{\omega}$ | Angular velocity vector |
| $\upsilon$ | Poisson's ratio |

*Subscripts*

| | |
|---|---|
| $d$ | Drag |
| $f$ | Fluid |
| $i$ | Particle $i$ |
| $j$ | Particle $j$ |
| $n$ | Normal |
| $p$ | Particle |
| $s$ | Solid |
| $t$ | Tangential |

## Data Availability Statement

Please contact the corresponding author for any requests for data, including code, from this work.

## References

[1] J. Cui, Q. Hou, and Y. Shen, "CFD-DEM study of coke combustion in the raceway cavity of an ironmaking blast furnace," *Powder Technology,* vol. 362, pp. 539-549, 2020.

[2] K. Kerst *et al.*, "CFD-DEM simulations of a fluidized bed crystallizer," *Chemical Engineering Science,* vol. 165, pp. 1-13, 2017.

[3] L. Lu, X. Gao, M. Shahnam, and W. A. Rogers, "Bridging particle and reactor scales in the simulation of biomass fast pyrolysis by coupling particle resolved simulation and coarse grained CFD-DEM," *Chemical Engineering Science,* vol. 216, p. 115471, 2020.

[4] S. Zhang, M. Zhao, W. Ge, and C. Liu, "Bimodal frequency distribution of granular discharge in 2D hoppers," *Chemical Engineering Science,* vol. 245, p. 116945, 2021.

[5] A. Kiani Moqadam, A. Bedram, and M. H. Hamedi, "A Novel Method (T-Junction with a Tilted Slat) for Controlling Breakup Volume Ratio of Droplets in Micro and Nanofluidic T-Junctions," (in en), *Journal of Applied Fluid Mechanics,* vol. 11, no. 5, pp. 1255-1265, 2018, doi: https://doi.org/10.29252/jafm.11.05.28598.

[6] P. A. Cundall and O. D. L. Strack, "A discrete numerical model for granular assemblies," *Géotechnique,* vol. 29, no. 1, pp. 47-65, 1979, doi: 10.1680/geot.1979.29.1.47.

[7] Y. Tsuji, T. Kawaguchi, and T. Tanaka, "Discrete particle simulation of two-dimensional fluidized bed," *Powder Technology,* vol. 77, no. 1, pp. 79-87, 1993/10/01/ 1993, doi: https://doi.org/10.1016/0032-5910(93)85010-7.

[8] B. Hoomans, J. Kuipers, W. J. Briels, and W. P. M. van Swaaij, "Discrete particle simulation of bubble and slug formation in a two-dimensional gas-fluidised bed: a hard-sphere approach," *Chemical Engineering Science,* vol. 51, no. 1, pp. 99-118, 1996.

[9] B. H. Xu and A. B. Yu, "Numerical simulation of the gas-solid flow in a fluidized bed by combining discrete particle method with computational fluid dynamics," *Chemical Engineering Science,* vol. 52, no. 16, pp. 2785-2809, 1997/08/01/ 1997, doi: https://doi.org/10.1016/S0009-2509(97)00081-X.

[10] R. Garg, J. Galvin, T. Li, and S. Pannala, "Open-source MFIX-DEM software for gas–solids flows: Part I—Verification studies," *Powder Technology,* vol. 220, pp. 122-137, 2012/04/01/ 2012, doi: https://doi.org/10.1016/j.powtec.2011.09.019.

[11] R. Garg, J. Galvin-Carney, T. Li, and S. Pannala, "Documentation of open-source MFIX–DEM software for gas-solids flows," *Tingwen Li Dr.,* 09/01 2012.

[12] C. Kloss, C. Goniva, G. Aichinger, and S. Pirker, "Comprehensive DEM-DPM-CFD simulations-model synthesis, experimental validation and scalability," in *Proceedings of the seventh international conference on CFD in the minerals and process industries, CSIRO, Melbourne, Australia*, 2009, pp. 9-11.

[13] C. Kloss, C. Goniva, A. Hager, S. Amberger, and S. Pirker, "Models, algorithms and validation for opensource DEM and CFD-DEM," *Pcfd,* vol. 12, p. 140, 2012.

[14] A. Kianimoqadam and J. L. Lapp, "Asynchronous GPU-based DEM solver embedded in commercial CFD software with polyhedral mesh support," *Powder Technology,* vol. 444, p. 120040, 2024/08/01/ 2024, doi: https://doi.org/10.1016/j.powtec.2024.120040.

[15] Y. He, F. Muller, A. Hassanpour, and A. E. Bayly, "A CPU-GPU cross-platform coupled CFD-DEM approach for complex particle-fluid flows," *Chemical Engineering Science,* vol. 223, p. 115712, 2020/09/21/ 2020, doi: https://doi.org/10.1016/j.ces.2020.115712.

[16] Z. Peng, E. Doroodchi, C. Luo, and B. Moghtaderi, "Influence of void fraction calculation on fidelity of CFD-DEM simulation of gas-solid bubbling fluidized beds," *AIChE Journal,* vol. 60, no. 6, pp. 2000-2018, 2014/06/01 2014, doi: https://doi.org/10.1002/aic.14421.

[17] A. Volk and U. Ghia, "Theoretical Analysis of Computational Fluid Dynamics–Discrete Element Method Mathematical Model Solution Change With Varying Computational Cell Size," *Journal of Fluids Engineering,* vol. 141, no. 9, 2019, doi: 10.1115/1.4042956.

[18] Z. Peng, B. Moghtaderi, and E. Doroodchi, "A modified direct method for void fraction calculation in CFD–DEM simulations," *Advanced Powder Technology,* vol. 27, no. 1, pp. 19-32, 2016/01/01/ 2016, doi: https://doi.org/10.1016/j.apt.2015.10.021.

[19] L. Wang, J. Ouyang, and C. Jiang, "Direct calculation of voidage in the fine-grid CFD–DEM simulation of fluidized beds with large particles," *Particuology,* vol. 40, pp. 23-33, 2018/10/01/ 2018, doi: https://doi.org/10.1016/j.partic.2017.11.010.

[20] D. A. Clarke, A. J. Sederman, L. F. Gladden, and D. J. Holland, "Investigation of Void Fraction Schemes for Use with CFD-DEM Simulations of Fluidized Beds," *Industrial & Engineering Chemistry Research,* vol. 57, no. 8, pp. 3002-3013, 2018/02/28 2018, doi: 10.1021/acs.iecr.7b04638.

[21] K. Takabatake and M. Sakai, "Flexible discretization technique for DEM-CFD simulations including thin walls," *Advanced Powder Technology,* vol. 31, no. 5, pp. 1825-1837, 2020/05/01/ 2020, doi: https://doi.org/10.1016/j.apt.2020.02.017.

[22] Z. Wang and M. Liu, "Semi-resolved CFD–DEM for thermal particulate flows with applications to fluidized beds," *International Journal of Heat and Mass Transfer,* vol. 159, p. 120150, 2020/10/01/ 2020, doi: https://doi.org/10.1016/j.ijheatmasstransfer.2020.120150.

[23] H. Xiao and J. Sun, "Algorithms in a robust hybrid CFD-DEM solver for particle-laden flows," *Communications in Computational Physics,* vol. 9, no. 2, pp. 297-323, 2011.

[24] R. Sun and H. Xiao, "Diffusion-based coarse graining in hybrid continuum–discrete solvers: Theoretical formulation and a priori tests," *International Journal of Multiphase Flow,* vol. 77, pp. 142-157, 2015/12/01/ 2015, doi: https://doi.org/10.1016/j.ijmultiphaseflow.2015.08.014.

[25] R. Sun and H. Xiao, "Diffusion-based coarse graining in hybrid continuum–discrete solvers: Applications in CFD–DEM," *International Journal of Multiphase Flow,* vol. 72, pp. 233-247, 2015/06/01/ 2015, doi: https://doi.org/10.1016/j.ijmultiphaseflow.2015.02.014.

[26] H. P. Zhu and A. B. Yu, "Averaging method of granular materials," *Physical Review E,* vol. 66, no. 2, p. 021302, 2002.

[27] A. Kianimoqadam and J. Lapp, "Calculating the view factor of randomly dispersed multi-sized particles using hybrid GRU-LSTM recurrent neural networks regression," *International Journal of Heat and Mass Transfer,* vol. 202, p. 123756, 2023/03/01/ 2023, doi: https://doi.org/10.1016/j.ijheatmasstransfer.2022.123756.

[28] Y. C. Zhou, B. D. Wright, R. Y. Yang, B. H. Xu, and A. B. Yu, "Rolling friction in the dynamic simulation of sandpile formation," *Physica A: Statistical Mechanics and its Applications,* vol. 269, no. 2, pp. 536-553, 1999/07/15/ 1999, doi: https://doi.org/10.1016/S0378-4371(99)00183-1.

[29] H. Hertz, "Ueber die Berührung fester elastischer Körper," 1882.

[30] R. D. Mindlin and H. Deresiewicz, "Elastic spheres in contact under varying oblique forces," 1953.

[31] T. B. Anderson and R. Jackson, "Fluid Mechanical Description of Fluidized Beds. Equations of Motion," *Industrial & Engineering Chemistry Fundamentals,* vol. 6, no. 4, pp. 527-539, 1967/11/01 1967, doi: 10.1021/i160024a007.

[32] R. Beetstra, M. A. van der Hoef, and J. A. M. Kuipers, "Drag force of intermediate Reynolds number flow past mono- and bidisperse arrays of spheres," *AIChE Journal,* https://doi.org/10.1002/aic.11065 vol. 53, no. 2, pp. 489-501, 2007/02/01 2007, doi: https://doi.org/10.1002/aic.11065.

[33] R. Beetstra, M. A. van der Hoef, and J. A. M. Kuipers, "Numerical study of segregation using a new drag force correlation for polydisperse systems derived from lattice-Boltzmann simulations," *Chemical Engineering Science,* vol. 62, no. 1, pp. 246-255, 2007/01/01/ 2007, doi: https://doi.org/10.1016/j.ces.2006.08.054.

[34] B. G. M. van Wachem, J. van der Schaaf, J. C. Schouten, R. Krishna, and C. M. van den Bleek, "Experimental validation of Lagrangian–Eulerian simulations of fluidized beds," *Powder*

*Technology,* vol. 116, no. 2, pp. 155-165, 2001/05/23/ 2001, doi: https://doi.org/10.1016/S0032-5910(00)00389-2.

[35] B. G. M. Van Wachem, J. C. Schouten, R. Krishna, and C. M. Van den Bleek, "Validation of the Eulerian simulated dynamic behaviour of gas–solid fluidised beds," *Chemical Engineering Science,* vol. 54, no. 13-14, pp. 2141-2149, 1999.